\documentclass[11pt]{article}
\usepackage{jheppub}

\usepackage{amsmath,graphicx,amssymb,bbm,subfigure,hyperref,array,slashed,cancel,physics,bbm}
\usepackage[all]{xy}
\usepackage{caption}
\usepackage{multirow}

\usepackage{cancel}

\newcommand\ignore[1]{}
\def\one{{\,\hbox{1\kern-.8mm l}}}

\def\Tr{{\rm Tr\, }}

\def\a{\alpha}\def\b{\beta}

\def\G{\Gamma}\def\e{\epsilon}

\def\m{\mu}
\def\n{\nu}

\def\r{\rho}\def\k{\kappa}
\def\s{\sigma}

\def\l{\lambda}

\def\d{\partial}

\def\ov{\overline}

\newcommand{\cH}{\mathcal H}

\newcommand{\cL}{\mathcal L}

\newcommand{\cR}{\mathcal R}

\newcommand{\bea}{\begin{eqnarray}}

\newcommand{\eea}{\end{eqnarray}}

\def\be{\begin{equation}}
\def\ee{\end{equation}}
\def\ba{\begin{eqnarray}}
\def\ea{\end{eqnarray}}

\def\k{\kappa}
\def\r{\rho}
\def\a{\alpha}

\def\b{\beta}

\def\G{\Gamma}

\def\e{\epsilon}

\def\th{\theta}

\def\m{\mu}
\def\n{\nu}
\def\om{\omega}
\def\Om{\Omega}
\def\l{\lambda}

\def\s{\sigma}

\def\cH{{\cal H}}

\def\cL{{\cal L}}

\def\cR{{\cal R}}

\def\qq{\qquad}

\def\IR{\relax{\rm I\kern-.18em R}}

\def\inv{^{\raise.0ex\hbox{${\scriptscriptstyle -}$}\kern-.05em 1}}

\def \ov {\over}

\title{Penrose limits of Abelian and non-Abelian T-duals of  $AdS_5\times S^5$ and their field theory duals}

\author{Georgios Itsios$^{1}$,}
\author{Horatiu Nastase$^{1}$,}
\author{Carlos N\'u\~nez$^{2}$,}
\author{Konstantinos Sfetsos$^{3,4}$}
\author{and Salom\'on Zacar\'ias$^{2}$}

\affiliation{$^1$ Instituto de F\'isica Te\'orica, UNESP-Universidade Estadual Paulista,
R. Dr. Bento T. Ferraz 271, Bl. II, Sao Paulo 01140-070, SP, Brazil}
\affiliation{$^2$ Department of Physics, Swansea University, Swansea SA2 8PP, United Kingdom}
\affiliation{$^3$ Department of Nuclear and Particle Physics, Faculty of Physics, National and Kapodistrian \\
University of Athens, Athens 15784, Greece}
\affiliation{$^4$ Theoretical Physics Department, CERN, CH-1211 Geneva 23, Switzerland}

\emailAdd{gitsios@gmail.com, nastase@ift.unesp.br, 
c.nunez@swansea.ac.uk, ksfetsos@phys.uoa.gr, salomon.zacarias@swansea.ac.uk} 

\abstract{We consider the backgrounds obtained by Abelian and non-Abelian
T-duality applied on $AdS_5\times S^5$. We study geodesics, 
calculate Penrose limits and 
find the associated plane-wave geometries. We quantise the weakly 
coupled type-IIA string theory on these backgrounds.
We study the BMN sector, finding  
operators that wrap the original quiver CFT. 
For the non-Abelian plane wave, we find a 'flow' in the frequencies. 
We report some progress to understand this, in terms of deconstruction of a higher dimensional field theory. 
We explore a relation with the plane-wave limit of the Janus solution, which we also provide. 
\\[10pt]
 } 

\subheader{}

\begin{document}
\def\Tr{{\textrm{Tr}}}


\hfill { CERN-TH-2017-249}


\maketitle

\newpage

\section{Introduction: Penrose limits,  pp-waves and  non-Abelian T-duality}

In the mid-seventies 't Hooft  suggested  
that there was a correspondence (a 'Duality') 
between string and gauge theories. Indeed, analyzing
the diagrammatic expansion of an $SU(N)$ gauge theory with 
arbitrary matter content, 't Hooft proposed that the relevant expansion parameter 
was $\lambda= g_{YM}^2 N$. The perturbation theory was written 
as a double-expansion in $\lambda$ and $\frac{1}{N^2}$ \cite{tHooft:1973alw}.
The correspondence was proposed to be with an unknown string theory with 
expansion parameter $g_s=\frac{1}{N}$ 
and world-sheet expansion parameter $\alpha'=\lambda^{-1/2}$.

The discovery of a precise duality between gauge theory 
and string theory---the Maldacena Conjecture 
\cite{Maldacena:1997re}, 
has allowed important progress in the stringy description of gauge theories. 
The string theoretic descriptions of several (confining or conformal) 
gauge theories with a large number of colors have since 
been found. In fact, backgrounds on which the dual 
type-II/M-theory should be formulated have been written for 
various gauge theories. Unfortunately, the string theory 
on these backgrounds is not soluble in general. 
Even when the masses of low-lying low-spin hadrons 
(or dimensions of special chiral operators in the conformal case) 
can be computed from supergravity, 
the full hadron spectrum (or spectrum of dimensions for 
non-chiral operators) requires calculating with the full string theory. 

It was the remarkable work of BMN \cite{Berenstein:2002jq}, that opened avenues to truly stringy calculations in gauge theories. 
Let us summarise the most salient features of the topic.
\subsection{Penrose limits and pp-waves}
Consider the 
$AdS_5\times S^5$ background, in global coordinates and for a space of radius $L$,
\begin{equation}
 \begin{aligned}
  ds^2 & = 4L^2 \big(-\cosh^2\rho \, dt^2+ d\rho^2+ \sinh^2\rho \, d\Omega^2_3 + d\alpha^2+\sin^2\alpha \, d\beta^2 \big)+
\\[5pt]
& +L^2 \cos^2\a \, \big( d\th^2+d\phi^2+d\psi^2 +2 \, \cos\th \, d\phi \, d\psi \big) \, ,
\\[5pt]
F_5 & = \frac{2}{g_s \, L} \, \big( 1 + *_{10} \big) \, \textrm{Vol}(AdS_5) \, ,\quad
L^4= \frac{\pi}{4} \, \a'^2 g_s \, {N} \, .
 \end{aligned}
\label{ads5xs5}
\end{equation}
Taking the Penrose limit along a geodesic sitting at $\rho=0$  and $\alpha=\frac{\pi}{2}$
leads to another maximally SUSY background. In fact, after defining 
\begin{eqnarray}
& & \rho=\frac{r}{L},\;\;\; \alpha=\frac{\pi}{2} +\frac{y}{L},\;\;\;  x^{+}=\frac{t+\beta}{2},\;\;\; x^-=L^2\frac{(t-\beta)}{2}\ , 
\end{eqnarray}
a plane wave space time arises when $L\to\infty$, 
and keeping only the leading order terms \cite{Blau:2002dy,Blau:2002mw}, we get
\begin{eqnarray}
& & ds^2=4 dx^+ dx^- - \mu^2(\vec{r}_4^2 +\vec{y}_4^2) (dx^+)^2 + d\vec{y}_4^2 + d\vec{r}_4^2\ ,
\nonumber\\
& & F_{+1234}=F_{+5678}=\mu\ .
\end{eqnarray}
The parameter $\mu$ was introduced by rescaling the $x^{\pm}$ coordinates. 
This Penrose limit on the gravity solution of (\ref{ads5xs5}), corresponds
on the field theory side to a restriction to operators having large $U(1)_R$ charge $J$, while simultaneously taking the large-$N$ limit.
This is called the 'BMN-sector' of ${\cal N}=4 $ Super-Yang-Mills. 
It comprises operators with dimension $\Delta$ and R-charge $J$ such that
\begin{eqnarray}
& &\frac{p^-}{\mu}= \Delta-J=\mathrm{fixed}\ ,\;\;\; \alpha' \mu p^+= \frac{\Delta+J}{\sqrt{4 g_{YM}^2 N }}=\mathrm{fixed} \ .
\nonumber\\
& & g_{YM}^2= \mathrm{fixed},\;\;\; \frac{J^2}{N}=\mathrm{fixed}\ ,\;\;\; N\to\infty\ ,\;\;\; J\to\infty\ .
\label{bmnscaling}
\end{eqnarray}
Hence, the full plane wave light-cone {\it string theory} 
Hamiltonian (denoted above by $p^-$) is equated with the difference between
the {\it field theory} dilatation and R-charge operators. 
Similarly, the spectrum of strings is related  to the spectrum of the dilatation operator, 
when restricted to the BMN sector.
Also, the BMN proposal in (\ref{bmnscaling}), 
states that the {\it string theory} 
Hilbert space (consisting on the direct sum of zero, one, two, etc, string states) 
equals the Hilbert space of ${\cal N}=4$ SYM generated by the 
BMN operators acting on the vacuum. In the same line, single string states map to single trace 
operators in the CFT. In particular the {\it string theory vacuum state} 
is in correspondence with a (suitably normalized) 
BMN {\it field theory} operator acting on the CFT vacuum state
\begin{eqnarray}
|0,p^+> \to {\cal N}\ \Tr(Z^J)|0>\ ,
\end{eqnarray}
where $Z$ is a chiral multiplet with R-charge $J[Z]=1$. 

For the {\it string theory } excitations, BMN proposed that one should consider {field theory} operators that are 'nearly BPS', 
that is, those with $\Delta-J\sim O(1)$. The chiral multiplets mix with the nearly BPS operators and the precise expression of the excited 
string states in terms of field theory states is more elaborated. See the review
\cite{Sadri:2003pr}  for its most updated version.

As we see from (\ref{bmnscaling}), the t' Hooft coupling $\lambda$ diverges. Perturbative calculations in the CFT are generically not trustable. 
However, for the case of the BMN operators, being 'nearly BPS', the
nice properties of the chiral primaries are inherited. 
The BMN proposal provided a first example of a 
full-fledged interacting string theory that calculates 
perturbative aspects of a gauge theory. Indeed, since the pp-waves are exact solutions
of the string theory to all orders in $\alpha'$ \cite{Horowitz:1989bv}, this
makes them the most suitable scenario to perform such computations.
For these reasons, the BMN proposal stands, 
sixteen years after it appeared, as one of the important 
developments in the area of gauge-strings duality.
This idea generated various interesting spin-offs, 
for example, the relation between AdS/CFT and integrability.

Now, we  summarise  recent 
progress in a different area of supergravity and string theory, namely 
non-Abelian T-duality. The bulk of this paper will study 
the synergy between this duality and the BMN-Penrose limits of the corresponding geometries in our continual effort to 
elucidate the nature of their field theoretical duals.  
\subsection{A quick review of non-Abelian T-duality}
Non-Abelian T-duality \cite{delaOssa:1992vci}, the generalization of the Abelian T-duality symmetry of string 
theory to non-Abelian isometry groups, is a transformation between world-sheet field theories. Its extension 
to all orders in $g_s$ and $\alpha'$ remains however a technically hard open problem \cite{Alvarez:1993qi}. 
As a result, non-Abelian T-duality does not stand as a String Theory duality symmetry, as its Abelian counterpart does.
{In addition, the transformation of the RR-flux fields under 
non-Abelian T-duality remained unknown for many years which made 
non-Abelian T-duality of limited 
use in relation with the AdS/CFT developments.}

{The authors of  \cite{Sfetsos:2010uq} reignited the interest 
in this transformation by extending it to include RR-fluxes and by
highlighting its potential powerful applications as 
a solution generating technique in supergravity. 
The  interesting synergy between the Maldacena conjecture 
and non-Abelian T-duality was first pointed out in 
\cite{Lozano:2012au}-\cite{Itsios:2013wd}
and  further exploited and developed in 
\cite{Lozano:2012au}-\cite{variosa5}. These works have widely applied 
non-Abelian T-duality to generate new AdS backgrounds of relevance in different contexts.
}

In the papers  \cite{Lozano:2016kum,completions},  the field 
theoretical interpretation of non-Abelian T-duality (in the context of Holography) was first addressed in detail. 
One outcome of these works is that non-Abelian T-duality changes the dual field theory. In fact, 
the new  backgrounds generated through non-Abelian T-duality have dual CFTs different from those dual to the original backgrounds. 
This is in accordance with that fact that, contrary to its Abelian counterpart, non-Abelian T-duality has not been proven to be 
a string theory symmetry. The results in 
 \cite{Lozano:2016kum,completions}, open up an exciting new way to generate new quantum field theories in the context of Holography.

The general idea of this paper is to consider 
both the Abelian and non-Abelian T-dual of $AdS_5\times S^5$,  
and study the Penrose limit for each of these geometries.
The associated field theory interpretation for these two T-dual backgrounds 
\cite{Lozano:2016kum},  will be used  in the study of the resulting pp-waves and the corresponding BMN operators.

Before moving into the discussion of these topics, we summarize 
the various results and achievements of the present work.
\subsection{Summary and plan of this paper}
As anticipated above, in this work we focus on the Penrose limits of both the Abelian and 
non-Abelian T-dual backgrounds of $AdS_5\times S^5$. 
We will study the BMN operators and the spectrum of the strings 
in each background. In more detail, the outline and salient achievements
of this paper are:
\begin{itemize}
\item{In Section \ref{section1xx}, we study the Abelian T-dual 
of the solution in (\ref{ads5xs5}). We consider various 
geodesics and their Penrose limit.
One of the pp-waves generated is especially interesting,
so that we study the quantization of the string theory in that background and compute the different eigen-frequencies. 
We  restrict the range of the conserved angular momentum quantity $J$ to 
avoid tachyons. We present an argument explaining why geodesics 
like the ones discussed  in relation to the Abelian T-dual background have the possibility of tachyonic states.}
\item{In Section \ref{section2xx}, we perform a very similar treatment to that 
in Section \ref{section1xx}, but now for non-Abelian T-duality. 
We study geodesics, the corresponding Penrose limits and also present 
the Brinkmann form of the resulting pp-wave. In addition, we perform the quantization 
of the string on this geometry. 
The eigen-frequencies are in this case depending 
on the light-cone time to be denoted by $u$ in what follows. 
This is an unexpected feature. Indeed, in spite of having an $AdS_5$ 
factor in our original geometry, we encounter a 'flow' in the string frequencies. 
In a precise sense, our system is explicitly realising the 'flows in pp-waves' 
that \cite{Corrado:2002wi, Gimon:2002sf} attempted 
to construct in the past. 

For small values of the light-cone time $u$, the eigen-frequencies (masses) become  constants. 
This regime of the pp-wave has a more canonical BMN interpretation. 
On the other hand, the large-$u$ behaviour is similar to the pp-wave 
of the Abelian T-dual, studied in Section \ref{section1xx}. }

\item{In Section \ref{section3xx}, we discuss the supersymmetry 
preserved by the geometries in Sections \ref{section1xx} and \ref{section2xx}, by carefully discussing the fermionic variations. Noticeably, there does not seem to be any enhancement of the half-supersymmetry preserved by all pp-wave solutions,
even at the 'ends of the flow'. }

\item{In Section \ref{sectionQFT}, following the  BMN proposal, 
we study
the field theoretical aspects of our new pp-wave backgrounds. We focus our attention mostly
on the Abelian pp-wave.

In terms of the parameters of the system $N$ and $k$ 
(the modding of the R-symmetry direction by $Z_k$), we
will show that our strings interact very weakly, 
with either $g_s\sim 1/k$ or $g_s\sim 1/N$ for the Abelian and non-Abelian 
T-dual pp waves respectively.

We  study the global symmetries and charges labelling the BMN operators and we 
write precise expressions for the vacuum and near BPS operators.
Our BMN-operators 'wrap' around the quiver field theory dual to the original 
field theory and  
have non-zero winding. We present an intuitive argument explaining 
this fact. Also, we argue that (in our scaling of the parameters) 
there is a mixing between various BMN operators, 
that corrects the eigen-frequencies from the naively expected values. 
Finally, we discuss the BMN-sector associated with 
the non-Abelian pp-wave.   In that case, we present arguments
showing that the non-Abelian T-dual of $AdS_5\times S^5$ is 
dual to a higher dimensional field theory. Both for small and large values of
a coordinate this reduces to a 4d CFT. This is reflected 
by the pp-wave and its
eigenfrequencies. }
\end{itemize}
Some conclusions and future directions of research are 
written in Section \ref{conclu}. Various appendices with interesting and important
technical details complement the presentation.

\section{Penrose limit of an Abelian T-dual of $AdS_5\times S^5$}\label{section1xx}
%
In this section we perform the Penrose limit on the Abelian T-dual 
of $AdS_5\times S^5$, and analyze closed 
string quantization in the resulting pp-wave background.

\subsection{Abelian T-dual solution}

We calculate the Abelian T-dual of $AdS_5\times S^5$, 
in a direction on $S^5$. Here we adopt the 
conventions of \cite{Lozano:2016kum}. 
More precisely, we consider the T-dual of $AdS_5\times S^5/\mathbb{Z}_k$.

After the Abelian  T-duality the coordinate acted upon by $\mathbb{Z}_k$
is renamed  $\tilde \psi$ and takes values in $[0,2\pi k]$. In order 
to restore the common factor of $L^2$ (the "radius" of the 
solution), we rescale $\tilde \psi$ by $L^2/\a'$,
\be
\label{rescc}
\tilde \psi=\frac{L^2}{\a'} \, \psi.
\ee
The new coordinate has generically small range,
\be
\psi=\frac{\a'}{L^2} \, \tilde{\psi} \, \in \,  \big[ 0,2\pi k\a'/L^2 \big]\;,
\label{psismall}
\ee
unless $k$ is of order $L^2/\a' = \sqrt{4\pi g_s N}$,
where $g_s$ is the string coupling in the type-IIB theory. 

In summary,  after these scalings,
the Abelian T-dual of the $AdS_5\times S^5$ 
solution along the $\psi$-direction is
\begin{equation}
 \begin{split}
& ds^2=  4\, L^2 \, ds^2(\textrm{AdS}_5)+4 \, L^2 \, d\Omega_2^2(\a,\b)+
\frac{L^2 d\psi^2}{\cos^2\a}+L^2\cos^2\a \, d\Omega^2_2
(\chi,\xi)\ ,
   \\[5pt]
  &  B_2  = L^2 \, \psi \, \sin\chi \, d\chi \wedge d\xi \ , \qquad 
   F_4 = \frac{8 \, L^4}{g_s\sqrt{\a'}} \, \cos^3 \a \, \sin \a \, \sin\chi \, d\a \wedge d\beta \wedge d\chi \wedge d\xi \ ,
   \\[5pt]
   &  e^{-2 \Phi} = \frac{L^2\cos^2 \a}{g_s^2\a'} \ ,
\end{split}
\label{maka1}
\end{equation}
where
\be
 \begin{split}
 & ds^2(\textrm{AdS}_5) =  - \cosh^2 r \, dt^2 + dr^2 + \sinh^2r \, d\Om^2_3 \ ,
   \\[5pt]
   & d\Omega_2^2(\a,\b)=d\a^2+\sin^2\a \, d\b^2\ , \quad d\Omega_2^2(\chi,\xi)=d\chi^2+\sin^2\chi \, d\xi^2\;,
 \end{split}
\ee

In the following, we analyze geodesics and Penrose limits in this background.

\subsection{Penrose limits}

A Penrose limit is a focusing on the region near 
a null geodesic. Denoting the affine parameter along the null geodesic by $\l$, the condition for a  geodesic to exist is 
\be
0=\frac{d^2x^\m}{d\lambda^2}+{\Gamma^\m}_{\n \r}\frac{dx^\n}{d\lambda}\frac{dx^\r}{d\lambda}=
\frac{d^2x^\m}{d\lambda^2}+\frac{1}{2}g^{\m \s} \big( \d_\n g_{\s \r}+\d_\r g_{\s \n}-\d_\s g_{\n \r} \big) \frac{dx^\n}{d\lambda}\frac{dx^\r}{d\lambda} \ , \quad \forall \m\ .
\label{geodxx1}
\ee
For motion (velocity) in the direction 
$x^\k$, we need to have no acceleration in the other directions, hence
\be
{\Gamma^\m}_{\k \k}=0\  \Longrightarrow\ g^{\m \n}\d_\n g_{\k \k}=\d^\m g_{\k \k}=0   \ ,\quad \forall\ \m
\ ,
\label{geodxx2}
\ee
where we have assumed that $x^\k$ corresponds to an isometric coordinate and that $g^{\m\n}$ is finite. 
Aside from this, we impose the condition for the geodesic to be null, $ds^2=0$.
For the metric (\ref{maka1}) it is natural to consider motion in the isometric directions $\xi$, $\b$ and $\psi$.

For motion along $\xi$, the conditions  (\ref{geodxx1})-(\ref{geodxx2}) give
\begin{equation}
 \begin{aligned}
g^{\chi\chi}\d_\chi g_{\xi\xi}=0 &\ \Rightarrow\  \sin\chi \, \cos\chi=0\ ,
\\[5pt]
g^{\a\a}\d_\a g_{\xi\xi}=0& \ \Rightarrow \ \cos\a \, \sin\a \, \sin^2\chi=0\ .
 \end{aligned}
\end{equation}
The above two conditions lead either to
$\chi=(0 , \, \pi/2 , \, \pi)$ and $\a= (0 , \, \pi/2 , \, \pi)$ or $\chi= (0 , \pi )$.
However, the values {$\chi = (0, \, \pi)$} are out of consideration, since otherwise the 
coordinate $\xi$, on which we take the limit, 
has a vanishing pre-factor. The same 
is true for $\a=\pi/2$. That means that we need to 
have two geodesics, one for $\a= 0$ and $\chi=\pi/2$ and one for  $\a= \pi$ and $\chi=\pi/2$. It turns out that both geodesics lead to the same pp-wave solution, thus in the following we are going to consider the first one of them.
Around the first geodesic, we make the expansions
\be
r=\frac{\bar r}{2L}\;,\;\;\;
\a=\frac{x}{2L}\;,\;\;\;
\psi=\frac{y}{L}\;,\;\;\;
\chi=\frac{\pi}{2}+{\frac{z}{L}}\;,\;\;
t=x^+\;,\;\;\;
{\xi}={2} \, x^+ +\frac{x^-}{L^2}\;,
\ee
while keeping $\b$ unchanged 
(we do not scale it). 
In order to keep finite $e^{-2\Phi}$ and $F_4$ we need 
to redefine the string coupling as 
\be
g_s=\frac{L}{\sqrt{\a'}}\tilde g_s.
\label{gsredefx}
\ee
The resulting pp wave metric,
\be
ds^2= 4 \, dx^+ dx^-  
+d\bar r^2+\bar r^2d\Omega_3^2+dx^2+x^2d\b^2+dz^2+dy^2 - \left(\bar{r}^2+x^2+ 4\, z^2\right)
(dx^+)^2\;,
\ee
is already in the Brinkmann form and has to be complemented with  the fields 
\begin{equation}
\label{hdhs}
 \begin{aligned}
B_2& = 2 \, y \, dz \wedge dx^+ \ ,\qquad e^{-2\Phi} = \frac{1}{\tilde g_s^2} \ ,
\\[5pt]
F_4& = \frac{4 \, x}{\tilde g_s} dx \wedge d\b \wedge dz\wedge dx^+.
 \end{aligned}
\end{equation}

For motion in $\b$, the condition  (\ref{geodxx1})-(\ref{geodxx2}) gives
\be\label{gbb}
g^{\a\a}
\d_\a g_{\b\b}=2\sin\a\cos\a =0\;,
\ee
which means that either {$\a= (0, \, \pi/2 , \, \pi)$}, with 
arbitrary values for $\psi=\psi_0, \chi=\chi_0,\xi=\xi_0$. 
For the geodesic with $\a=\pi/2$ and $\psi=\psi_0$, 
$\chi=\chi_0$, $\xi=\xi_0$, we expand the coordinates as
\be
\a=\frac{\pi}{2}+{\frac{y}{L}} \;,\;\;\;
r=\frac{\bar r}{2L} \;,\;\;\;
t=x^+ \;,\;\;\;
\b=x^++\frac{x^-}{{2} \, L^2} \ ,
\label{betappxx}\ee
and we keep $(\chi,\xi)$ unchanged. We also keep the original coordinate $\tilde \psi$ unchanged, 
i.e. we undo the rescaling (\ref{rescc}).
Then we obtain the  metric
{
\be
ds^2= 4 \, dx^+dx^-+d\bar r^2+\bar r^2d\Omega_3^2+4 \, dy^2+y^2d\Omega^2_2(\chi,\xi)+\frac{\a'^2d\tilde\psi^2}{y^2}- (\bar r^2+ 4\, y^2)
(dx^+)^2\;,
\label{nopp1}
\ee
complemented by the fields
\begin{equation}
 \begin{aligned}
B_2&= \a' \tilde \psi \sin \chi \, d\chi \wedge d\xi \ ,\qq   e^{-2\phi}=\frac{y^2}{\a' g_s^2} \ ,
  \\[5pt]
F_4&= \frac{8 \, y^3}{g_s\sqrt{\a'}}\sin \chi \, dx^+ \wedge dy \wedge d\chi \wedge d\xi \ .
 \end{aligned}
\end{equation}
}
As we see, in this limit the original  coordinate $\tilde \psi$ (before the rescaling) and $g_s$ were the
appropriate quantities to use. We have checked that the metric (\ref{nopp1}) 
has non-vanishing scalar curvature and therefore it is not of the  plane wave form. 
Since the null geodesic sits at a singular position ($\a=\pi/2$) 
in the background, in the exact Penrose limit we don't obtain a pp wave form, though by a coordinate
transformation of (\ref{maka1}) (that would become singular in the exact Penrose limit), it might be possible to do so and thus respect Penrose's 
construction.

Whilst it would be interesting to consider the field theory 
interpretation of these two Penrose limits, 
we will be more interested  in understanding the 
effect of T-duality through the Penrose limit. 
With this purpose,  it is useful to consider the Penrose limit 
for motion along the $\psi$-direction. In this case, the geodesic 
condition in (\ref{geodxx1})-(\ref{geodxx2}) is just
\be
g^{\a\a} \, 
\d_\a g_{\psi\psi}=0 \ \Rightarrow \ \frac{\sin \a}{\cos^3\a}=0\;,
\ee
leading to $\a=0$. 
However, we will see that moving just along 
the $\psi$ direction involves some pathologies 
for the string propagation. We consider instead the combined 
motion on both the  $\psi$ and $\xi$ directions.

\subsection{Penrose limit in $\psi$ and $\xi$ and pp wave}

For motion along the $\psi$ and {$\xi$} directions, 
we must satisfy the geodesic conditions. 
This implies $\a=0$ and $\chi=\pi/2$. 
We can then construct a Lagrangean for 
a (massless) particle moving on a geodesic as above,
\be
{\cal L} = {\frac{1}{2}} \, g_{\mu\nu}\dot X^\mu \dot X^\nu\;,
\ee
where the dot refers to differentiation with respect to the 
affine parameter $u$ along the geodesic.
We will obtain  the geodesic needed and using  
an expansion around it, the corresponding pp-wave.
The Lagrangian is given by
\begin{equation}
\cL =\frac{L^2}{2} \, ( -4 \, \dot{t}^2 + \dot{\psi}^2 + \dot{\xi}^2 ) \ ,
\end{equation}
which is independent of $(t,\psi,\xi)$, since the coordinates are cyclic. This
leads to the conservation of their conjugate momenta
\begin{equation}
\label{papax}
\frac{\partial \cL}{\partial \dot{t}} = -4 L^2 \dot t = -L^2 \ , \qquad \frac{\partial \cL}{\partial \dot{\xi}} 
 = L^2 \dot \xi  = - J L^2 \ , 
 \qquad \frac{\partial \cL}{\partial \dot{\psi}}=  L^2\dot \psi  = \textrm{const} \ ,
\end{equation}
where we have appropriately fixed the energy $L^2p_t$ which in turn fixes the relation between time and the 
affine parameter $u$. The conserved quantity associated with the cyclic variable $\xi$ is parametrized by $J$ and that for 
$\psi$ is  determined from the fact that the geodesic must be null, i.e. 
${\cal L}=0$. This gives
\begin{equation}
 \dot{\psi}^2 = \frac{1}{4} \, \big(   1 - 4 \, J^2  \big) \quad \Longrightarrow\quad
  \psi = \frac{\sqrt{1 - 4 \, J^2}}{2} \, u\ , 
\label{papaxx}
\end{equation}
where in the solution  we have ignored an additive integration constant. 
We note that for the reality of the solution 
we need $J^2\leqslant 1/4$, or (ignoring the 
trivial sign of $J$ that just defines the direction of motion in $\xi$)
\be
\label{injj}
0\leqslant J\leqslant \frac{1}{2}\ .
\ee
We also note that $\psi\sim {\cal O}(1)$ 
means $\tilde \psi\sim {\cal O}(L^2/\a')$, 
but we also need $\tilde \psi\leq 2\pi k$. That means that 
in order to have $\psi$ of order one 
(as suggested by the above formulas), we need to have
\be
\frac{L^2}{\a'}=\sqrt{4\pi g_s^BN}\sim k\;,
\ee
where $g_s^B$ is the string coupling in the type IIB theory 
(before the T-duality). 

In (\ref{papaxx}) above, $u$ is the affine parameter along 
the null geodesic, i.e., the ``lightcone time" (usually denoted by $x^+$). To define the Penrose limit 
around $r=0, \;\,\a=0,\; \,\chi=\pi/2$, we  expand 
\begin{equation}
 r = \frac{\bar{r}}{2 \, L} \ , 
\qquad \a = \frac{x}{2 \, L} \ , \qquad \chi = \frac{\pi}{2} + \frac{z}{L} .
\label{pappaxx}
\end{equation}
In principle there are several ways to take the Penrose limit 
(see Appendix \ref{appendixA} for details), 
but the more physical one is where we 
do not  have a compact time \cite{Russo:2002rq}. 

%
Hence, the coordinate $t$ must be  proportional to the lightcone time $u$
as we  found above. We propose an expansion near 
the non-trivial geodesic  in the $\psi$ direction, with a component in the 
$\xi$ direction.  That restricts us to the ansatz
\begin{equation}
 \begin{split}
  & dt = c_1 \, du \ ,
  \\[5pt]
  & d\xi = c_2 \, du + c_3 \, \frac{dw}{L} \ ,
  \\[5pt]
  & d\psi = c_4 \, du + c_5 \, \frac{dw}{L} + c_6 \, \frac{dv}{L^2} \ ,
 \end{split}
 \label{consts}
\end{equation}
where $c_i \ , i = 1 , \ldots , 6$ are constants. 
Moreover, the leading terms $c_1,c_2, c_4$ 
must be the ones from the null geodesic above 
\be
c_1=\frac{1}{4} \ , \quad c_2=- J  \ , \quad c_4 =  \frac{1}{2}\sqrt{1-4J^2}.
\ee
This ensures cancellation of the ${\cal O}(L^2)$ terms in the would be plane wave metric, 
when $L\rightarrow \infty$ as needed for the Penrose limit. 
For the cancellation of the ${\cal O}(L)$ terms, 
we obtain $c_2 c_3+ c_4 c_5=0$. 
Finally, demanding that the coefficient of $dw^2$ term is normalized to unity
we obtain $c_3^2+c_5^2=1$ and from the condition to have a $2 \,du \, dv$ 
term we get $c_4 c_6=1$. These conditions are solved by
\begin{equation}
 c_3 = \sqrt{1-4 J^2}\ ,\qq  c_5 = 2 J  \ , \qq c_6 = \frac{2}{\sqrt{1-4 J^2}} \ .
\end{equation}
We then obtain a plane wave solution in Brinkmann coordinates, with
\begin{equation}
 ds^2 = 2 \, du \, dv + d\bar{r}^2 + \bar{r}^2 \, d\Om^2_3 + dz^2 + dx^2 + x^2 \, d\b^2 + dw^2 
 - \Big[  \frac{\bar{r}^2}{16} +{8J^2 -1\ov 16} \, x^2 + J^2 \, z^2  \Big] \, du^2
\label{abppxx1}
\end{equation}
and
\begin{equation}
 \begin{aligned}
 e^{2 \, \Phi} = g_s^2\frac{\a'}{L^2}\equiv \tilde g_s^2 \ , 
 \qquad B_2 = \frac{u}{2} \, dz \wedge dw \ ,\qquad F_4 = \frac{2 \, J \, x}{\tilde g_s} \, du \wedge dz \wedge dx \wedge d\b \ .
 \end{aligned}
\label{abppxx2}
\end{equation}
Note that the coefficient of the $x^2 (du)^2$ term may be negative for low enough values for $J$ which implies the appearance 
of tachyonic modes. Combining with (\ref{injj}) we obtain the condition for physical string propagation on this background 
\be
\frac{1}{2\sqrt{2}}\leqslant J\leqslant \frac{1}{2}\;,
\label{restrictxx}
\ee
which excludes $J=0$, as advertised. We will come back to this in the next subsection.
Technically, the origin of the potential tachyonic mode leading to (\ref{restrictxx}) is the difference
of the $AdS_5\times S^5$ in (\ref{ads5xs5}) and the Abelian T-dual metric in (\ref{maka1}) 
in the $\psi$ direction. Using the scaling in (\ref{pappaxx}), we find that
\bea
& & ds^2_{S^5}\sim \cdots + \cos^2\alpha 
d\psi^2\to \cdots +\left(1-\frac{x^2}{4L^2}\right)d\psi^2,\nonumber\\[5pt]
& & ds^2_{\rm T-dual}\sim \cdots + \frac{1}{\cos^2\alpha} 
d\psi^2\to \cdots +\left(1+\frac{x^2}{4L^2}\right)d\psi^2.
\label{mecxx}
\eea
It is the sign difference in the above expansions that introduces the 
sign flip responsible for the potential tachyonic mode.

Subsequently we study string quantization in the background given by (\ref{abppxx1})-(\ref{abppxx2}).

\subsection{Closed string quantization on the pp wave}

The action for a string moving in a space with metric $G_{\m \n}$, 
in the presence of the background fields $B_{\m \n}$ and $\Phi$ is
\begin{equation}
\label{BosonicAction}
  S = - \frac{1}{4 \, \pi \, \a'} \int d\tau \, d\s \Big[  \sqrt{g} \, g^{\a \b} \, G_{\m \n} \, \partial_\a X^\mu \, \partial_\b X^\n + \e^{\a \b} \, 
  B_{\m \n} \, \partial_\a X^\mu \, \partial_\b X^\n + \a' \, \sqrt{g} \, \cR^{(2)} \, \Phi    \Big] \ ,
\end{equation}
where the antisymmetric tensor density, 
the combination $\e^{\a \b}/\sqrt{g}$, transforms as a tensor and
$\e^{\a \b}$ has non-vanishing components $\e^{\tau \s} = - \e^{\s \tau} = 1$.
We fix the Weyl symmetry such that
\begin{equation}
\label{WeylFix}
 \sqrt{-g} \, g^{\a \b} =  \eta^{\a \b} \ , \qquad - \eta_{\tau \tau} = \eta_{\s \s} = 1 \ .
\end{equation}
We will use the notation
\begin{equation}
\label{coordCorresp}
 \underbrace{U , \, V}_{u , \, v} , \, \underbrace{X^1 , \, X^2 , \,  X^3 , \, X^4}_{\bar{r} , \, \Om_3} , \, \underbrace{X^5 , \, X^6}_{x , \, \b} , \,  
 \underbrace{X^7 , \, X^8}_{z , \, w} \ . 
\end{equation}
In order to fix the residual diffeomorphism invariance we take
\begin{equation}
\label{DiffeoFix}
 U = \tau \ .
\end{equation}
Then, for the plane wave in (\ref{abppxx1})-(\ref{abppxx2}), we obtain that 
\begin{equation}
 \begin{aligned}
  S = & - \frac{1}{4 \, \pi \, \a'} \int d\tau \, d\s \Bigg[   \partial X^i \cdot \partial X^i 
  + \frac{\big(  X^1  \big)^2 + \big(  X^2  \big)^2 + \big(  X^3  \big)^2 + \big(  X^4  \big)^2}{16} 
  \\[5pt]
  & + \frac{\big(  X^5  \big)^2 + \big(  X^6  \big)^2}{16} \,  (8 J^2 - 1) + J^2 \big(  X^7  \big)^2
  - \Big(  \tilde{\k}_1 \, X^7 \, \partial_\s X^8 - \tilde{\k}_2 \, X^8 \, \partial_\s X^7   \Big)  \Bigg] \ ,
 \end{aligned}
\end{equation}  
where the inner product is defined using the metric $\eta_{\a \b}$ and 
we have considered the following gauge choice for the NS two-form,
\begin{equation}
   B_2= \frac{1}{2} \, \Big(  \tilde{\kappa}_2 \, w \, du \wedge dz -   \tilde{\kappa}_1 \, z \, du \wedge dw  \Big) \ , \qquad \tilde{\kappa}_1 + \tilde{\kappa}_2 = 1 \ .
\end{equation}
The equations of motion for the scalars in the above action read
\begin{equation}
 \begin{split}
  & \Box X^i - \frac{1}{16} \, X^i = 0 \ , \qquad i = 1, \ldots , 4 \ ,
  \\[5pt]
  & \Box X^i - \frac{8 \, J^2 - 1}{16} \, X^i = 0 \ , \qquad i = 5, 6 \ ,
  \\[5pt]
  & \Box X^7 - J^2 \, X^7 + \frac{ 1 }{2} \, \partial_\s X^8 = 0 \ ,
  \\[5pt]
  & \Box X^8 - \frac{ 1 }{2} \, \partial_\s X^7 = 0 \ .
 \end{split}
\end{equation}
The corresponding boundary conditions are
\begin{equation}
 \begin{split}
  & \big( \partial_\s X^i \big) \, \delta X^i \Big|_{\s = 0}^{\s = 2 \pi} = 0 \ , \qquad i = 1 , \ldots , 6 \ ,
  \\[5pt]
  & \Big(   \partial_\s X^7 +  \frac{ \tilde{\kappa}_2 }{2} \, X^8  \Big) \, \delta X^7 \Big|_{\s = 0}^{\s = 2 \pi} = 0 \ ,
  \\[5pt]
  & \Big(   \partial_\s X^8 - \frac{ \tilde{\kappa}_1 }{2} \, X^7  \Big) \, \delta X^8 \Big|_{\s = 0}^{\s = 2 \pi} = 0 \ .
 \end{split}
 \label{bdconds}
\end{equation}
Using the above equations of motion and an ansatz of the form $X^i \sim e^{-i \om t + i n \s}$, we obtain the  
frequencies for the oscillators
\begin{equation}
 \begin{split}
  & \om^2_{n, i} = n^2 + \frac{1}{16} \ , \qquad i = 1 , \ldots , 4 \ ,
  \\[5pt]
  & \om^2_{n, i} = n^2 + \frac{8 \, J^2 - 1}{16} \ , \qquad i = 5, 6 \ ,
  \\[5pt]
  & \om^2_{n, \pm} = n^2 + \frac{J^2}{2} \pm \frac{1}{2} \, \sqrt{n^2 + J^4} \ .
 \end{split}
\label{freqabx}
\end{equation}
As we anticipated, the frequencies for $i=5,6$ are not  tachyonic if the restriction (\ref{restrictxx}) is obeyed.

\section{Penrose limit of the non-Abelian T-dual  of $AdS_5\times S^5$}\label{section2xx}

In this section we study  the non-Abelian case by repeating the same steps as for the Abelian one of the previous section. Namely,  we write the T-dual solution, take Penrose limits and quantize closed strings in the resulting background. 

\subsection{Non-Abelian T-dual solution}

We consider the non-Abelian T-dual along an $SU(2)$ isometry 
of the same $AdS_5\times S^5/\mathbb{Z}_k$ background. 
After T-duality the coordinates are denoted by 
$\tilde \rho\in \big[ 0, \, 2\pi k \big]$ and the two angles $(\chi,\xi)$.\footnote{In principle, there is not a restriction 
on the range of the 'radial' dual coordinate. 
The ideas in \cite{Lozano:2016kum} 
indicate that the 'radial' coordinate divides naturally 
in intervals of size $ \pi$.}

The non-Abelian T-dual solution has NS-NS sector \cite{Sfetsos:2010uq}
\begin{equation}
\label{NATDsolNSor}
 \begin{aligned}
   ds^2 & = 4\, L^2 \, ds^2(\textrm{AdS}_5)+4 \, L^2 \, d\Omega_2^2(\a,\b)
 + \frac{\a'^{2}d\tilde\r^2}{L^2 \, \cos^2 \a}  +\frac{\a'^2L^2\tilde\rho^2\cos^2\a}{\a'^2\tilde\rho^2
+L^4\cos^4\a}d\Omega^2_2(\chi,\xi)\; ,
\\[5pt]
   B_2 & = \frac{\a'^3 \tilde \r^3}{\a'^2 \tilde \r^2 + L^4 \cos^4 \a} \sin\chi \, d\chi \wedge d\xi \ , 
   \quad e^{-2 \Phi} = \frac{L^2 \cos^2 \a}{g_s^2 \, \a'^3} \big(   \a'^2 \tilde \r^2 + L^4 \cos^4 \a  \big) \ 
 \end{aligned}
\end{equation}
and Ramond fields
\bea
 F_2& =& \frac{8 \, L^4}{g_s \, \a'^{3/2}} \sin\a \, \cos^3 \a \, d\a \wedge d\b \ , \cr
 F_4 &=& \frac{8 \, \a'^{3/2} \, L^4}{g_s}\frac{\tilde \r^3 \, \cos^3\a}{\a'^2 \tilde \r^2 
 + L^4 \cos^4 \a} \, \sin\alpha \, \sin\chi \, d\alpha \wedge d\beta \wedge d\chi \wedge d\xi \ .\label{NATDsolRR}
\eea
\subsection{Penrose limits}
We start by discussing various possible geodesics.
Considering the metric above
we impose the condition in (\ref{geodxx2}), 
for a geodesic moving in $\b$ and we find the same constraint as in the Abelian case, namely (\ref{gbb}).
As in that case only the case $\a=\pi/2$ is viable with arbitrary values for 
$\rho$, $\chi$ and $\xi$.
Then, we propose the expansion  (\ref{betappxx})
keeping $\chi,\xi$ and $\tilde \rho$ unchanged. This leads to the metric
\begin{equation}
\label{nopp2} 
 \begin{aligned}
   ds^2 & = 4 \, dx^{+} \, dx^{-} + d\bar{r}^2 + \bar{r}^2 \, d\Om^2_3 + dy^2 -  (  \vec{x}^2_4 + y^2) \,
 dx^{+2} 
\\[5pt]
   & \quad +4 \frac{\a'^2}{y^2} \, d\tilde\r^2 + \frac{4 \, \a'^2 \tilde\r^2 y^2}{16 \, \a'^2 \tilde\r^2 + y^4}d\Omega^2_2(\chi,\xi) \ ,
 \end{aligned}
\end{equation}
where $\vec{x}^2_4$ is the radial distance squared in the 
four-dimensional space spanned by the spherical 
coordinates $\bar{r} , \, \Om_3$, so $\vec{x}^2_4 = \bar{r}^2$.
The NS-NS fields become, in the limit of large $L$,
\begin{equation}
 B_2 = \frac{16 \, \a'^3\tilde \r^3}{ 16 \, \a'^2 \tilde\r^2 + y^4 } \, \sin \chi \, d\chi \wedge d\xi \ , 
 \qquad e^{-2\Phi} =g_s^{-2}\   \frac{y^2 \big(  16 \, \a'^2 \tilde\r^2 + y^4  \big)}{64\a'^3} \;,
\end{equation}
and the RR fields are
\begin{equation}
 g_s \, F_2 = \frac{y^3}{2 \, \a'^{3/2}} dx^{+} \wedge dy \ , \qquad g_s \, F_4 = 
 \frac{8 \, \a'^{3/2} y^3 \tilde\r^3}{16 \, \a'^2 \tilde\r^2 + y^4} \, 
 \sin \chi \, dx^{+} \wedge dy \wedge d\chi \wedge d\xi  \ .
\end{equation}
 We have checked that the metric (\ref{nopp2}), similar 
to (\ref{nopp1}), 
has non-vanishing scalar curvature and therefore it is not in a plane wave form. 
We will not discuss this background in the following.

On the other hand, if 
we consider a geodesic along $\xi$, the conditions in (\ref{geodxx2}) give
\bea
g^{\tilde \rho\tilde \rho}
\d_{\tilde\rho} g_{\xi\xi}=0 \quad &\Rightarrow & \quad\frac{2 \, L^8 \, \tilde \rho \, \sin^2\chi \, \cos^8\a}{\a'^2\tilde \rho^2+L^4 \cos^4\a}=0 \ ,
\nonumber\\[5pt]
g^{\chi\chi}\d_\chi g_{\xi\xi}=0 \quad &\Rightarrow & \quad 2 \, \sin\chi \, \cos\chi=0 \ ,\label{condits}
\\[5pt]
g^{\a\a}\d_\a g_{\xi\xi}=0 \quad & \Rightarrow & \quad 2 \,  \a'^2 \, \tilde\rho^2 \, \cos^3\a \, \sin\a \, \sin^2\chi \,\frac{\a'^2\tilde\rho^2-L^4\cos^2\a}{(\a'^2\tilde\rho^2+L^4\cos^4\a)^2}=0 \ ,
\nonumber
\eea
which leads to the restrictions
\begin{itemize}
\item
$\tilde\rho=0$ or $\a=\pi/2$\ ,

\item
$\chi = (0 , \pi/2 , \pi)$\ ,

\item
$\a = (0 , \pi/2 , \pi)$ or $\chi = (0, \pi)$ or $L^4\cos^4\a=\a'^2\tilde\rho^2$, 

\end{itemize}
corresponding to the three above restrictions.
As in the Abelian case, the nonvanishing of the coefficient for $d\xi^2$ eliminates the possibilities of having $\chi=(0, \pi) $ and $\a=\pi/2$, but now it also eliminates $\tilde \rho=0$, 
which means that the first condition does not have a solution. Indeed,  in this case, unlike the Abelian background, 
motion in the $\xi$ direction alone does not correspond to a geodesic.

\subsection{Penrose limit in $\rho$ and $\xi$}

We may consider a combined motion in $\tilde \rho $ and $\xi$, in which case we can find a geodesic. 
In this case it is convenient to first rescale the coordinate $\tilde{\rho}$, analogously to (\ref{rescc}) in order 
to restore the common $L^2$ factor to the metric
\be
\tilde \rho =\frac{L^2}{\a'}\rho\ .
\label{rhosmall}
\ee
Hence, we obtain a solution that takes parametrically small values for $\rho$ unless $k$ is comparable to $L^2/\a'$. 
In addition we rescale the string coupling as
\be
\tilde{\tilde g}_s=g_s\frac{\a'^{3/2}}{L^3}\ .
\label{gsredefxx}
\ee
Consider the null geodesic for motion in $\rho$ and $\xi$, at $\chi=\pi/2$, $\a=0$, which would solve the last two conditions in (\ref{condits}), 
and $r=0$. 
The non-Abelian T-dual solution (\ref{NATDsolNS}) in the rescaled coordinate $\rho$, and in 
terms of the rescaled coupling reads 
\begin{equation}
\label{NATDsolNS}
 \begin{aligned}
   L^{-2} ds^2 & = 4 \, ds^2(\textrm{AdS}_5)+4 \, d\Omega_2^2(\a,\b)
  + \frac{d\r^2}{ \cos^2 \a} + \frac{\r^2  \cos^2 \a }{ \r^2 +  \cos^4 \a} 
   d\Omega_2^2(\chi,\xi) \ ,
\\[5pt]
   B_2 & = \frac{L^2  \r^3}{  \r^2 + \cos^4 \a} \sin\chi \, d\chi \wedge d\xi \ , 
   \qquad e^{-2 \Phi} = \frac{ \cos^2 \a}{\tilde{\tilde g}_s^2} \big(    \r^2 +  \cos^4 \a  \big) \;,
 \end{aligned}
\end{equation}
with the R-R fields 
\begin{equation}
\label{NATDsolRR}
 F_2 = \frac{8 \, L}{\tilde{\tilde g}_s} \sin\a \, \cos^3 \a \, d\a \wedge d\b \ , \qquad F_4 = \frac{8 \, L^3}{\tilde{\tilde{g}}_s}\frac{\rho^3 \, \cos^3\alpha}{\rho^2 + \cos^4\alpha} \, \sin\alpha \, \sin\chi \, d\alpha \wedge d\beta \wedge d\chi \wedge d\xi \ .
\end{equation}
The Lagrangean for a particle moving on a null geodesic reads
\begin{equation}
 L^{-2} \, \cL = {\frac{1}{2}} \Big(  -4 \, \dot{t}^2 + \dot{\r}^2 + \frac{\r^2}{\r^2 + 1} \, \dot{\xi}^2  \Big) \ ,
\end{equation}
where as before,  the dot indicates a derivative with respect 
to the affine parameter $u$. 
The Lagrangean is independent on $t$ and $\xi$ (these coordinates are cyclic), which means that their conjugate momenta are 
conserved, 
\begin{equation}
 \frac{\partial \cL}{\partial \dot{t}} = p_t = \textrm{const} , \qquad  \frac{\partial \cL}{\partial \dot{\xi}} = p_{\xi} = \textrm{const} \ .
\end{equation}
As in the Abelian case, we 
fix the energy and define the angular momentum 
in the $\xi$ direction by the conditions
\begin{equation}
L^{-2} \, p_t= -4 \, \dot{t} = -1 \ ,  \qquad L^{-2} \, p_\xi= \frac{\r^2}{\r^2 + 1} \dot{\xi} = - J \ .
\end{equation}
Then $\dot \rho$ is determined from the condition that 
the geodesic be null, ${\cal L}=0$, giving 
\begin{equation}
\label{eqrho}
 \dot{\r}^2 = \frac{1}{4} - \frac{\r^2 + 1}{\r^2} \, J^2 \ .
\end{equation}
This has solution
\begin{equation}
\label{solr}
 \r^2 = \frac{ \big(  1-4 J^2  \big)^2 (c_\r \pm u)^2 + 16 \, J^2}{4 \, (1-4 \, J^2)} \ , \qquad c_\r = \textrm{const} \ .
\end{equation}
As in the Abelian case, the affine parameter $u$ is now 
the lightcone time (usually denoted by $x^+$). 

Note that, for the right hand side of the equation to be positive, we need that $J^2\leqslant 1/4$ and, since $J\rightarrow -J$ is an
irrelevant choice of direction of motion on $\xi$, we have
\be
0\leqslant J\leqslant 1/2 \ .
\ee
For the same reason, $\rho$ is restricted to be 
\be
\rho^2\geqslant \frac{4J^2}{1-4J^2},
\ee
which in terms of the original coordinate $\tilde \rho$ reads
\be
\tilde \rho \geqslant \frac{L^2}{\a'}\frac{2J}{\sqrt{1-4J^2}}\;.
\ee
This must fit inside the interval $ \big[ 0, \, 2\pi k \big]$, which means that we need $k\sim L^2/\a'$. 

To define the Penrose limit, we expand around the null geodesic considered above, 
\begin{equation}
 r = \frac{\bar{r}}{2 L} \ , \qquad \a = \frac{x}{2 L} \ , \qquad \chi = \frac{\pi}{2} + \frac{z}{2 L}\;,
\end{equation}
and we set
\begin{equation}
\label{CoordTransModifiedFixed}
 \begin{aligned}
   & dt = c_1 \, du \ ,
  \\[5pt]
  & d\xi = c_2 \, du + \big(  1 + 4 \, J \, c_2  \big) \, \frac{dw}{L} +\Big(   c_2 - \frac{3}{4 \, J}  \Big) \, \frac{dv}{L^2} \ ,
  \\[5pt]
  & d\rho = c_3 \, \Big(  du + \frac{4 \, J }{L} \, dw +  \frac{dv}{L^2}  \Big) \;,
 \end{aligned}
\end{equation}
with
\be
c_1=\frac{1}{4} \, ,
\qquad
c_2= - \frac{\rho^2+1}{\rho^2}J \, ,
\qquad
c_3=\sqrt{\frac{1}{4}-\frac{1+\rho^2}{\rho^2}J^2}.
\ee
%
%
%

Substituting everything in the metric, we obtain the pp wave
\begin{equation}
\label{Metricpprxi}
 \begin{aligned}
   ds^2_{pp} & = 2 \, du \, dv + d\bar{r}^2 + \bar{r}^2 \, d\Om^2_3 + dx^2 + x^2 \, d\b^2 + \frac{\r^2}{\r^2 + 1} \, dz^2 
   + \Bigg[   \frac{\r^2}{\r^2 + 1} - 4 J^2  \Bigg] \, dw^2
   \\[5pt]
   & - \Bigg[   \frac{\bar{r}^2}{16} + \frac{x^2}{16} \,  (8 J^2 - 1) + \frac{\r^2 + 1}{\r^2} J^2 z^2  \Bigg] \, du^2 \ .
 \end{aligned}
\end{equation}
At this point, we can already observe that, unless $J^2\geq 1/8$, the $x^2 \, du^2$ term will have the wrong sign, which will lead 
to tachyonic modes for closed string propagation in this background. We will explore this in more detail in the next subsection, but for now
we just point out that the consistent range of $J$ is
\be
\frac{1}{2\sqrt{2}}\leqslant J\leqslant \frac{1}{2}\  ,
\ee
which is the same as \eqref{restrictxx}.
In particular, we see that $J=0$, i.e. motion solely along $\rho$, is not consistent, as anticipated above.

The field strength of the Neveu-Schwarz B-field in the Penrose limit becomes
\be
H_3 = d\Bigg[\frac{\r^3}{\r^2 + 1} \, dz \wedge dw \Bigg] \;,
\ee
which tells us that the B-field is
\begin{equation}
\label{NS2formrxi}
 B_2 = \frac{\r^3}{\r^2 + 1} \, dz \wedge dw \ .
\end{equation}
The dilaton in the limit becomes simply
\begin{equation}
\label{DilatonppNATD}
 e^{-2 \Phi} = \frac{\r^2 + 1}{\tilde{\tilde g}_s^2} \ .
\end{equation}
The Ramond fields in the $L\rightarrow \infty$ limit are
\begin{equation}
 F_2 = 0 \ , \qquad F_4 = - \frac{2 \, J \, x \, \r}{\tilde{\tilde g}_s} \, dz \wedge du \wedge dx \wedge d\b \ .
\end{equation}
The pp wave obtained above is not yet in Brinkmann coordinates, 
which is what we would need for a simple string quantization. 
Below, we discuss the coordinate change that brings the 
background to the desired form.
\subsubsection{Brinkmann form}
In order to bring the metric into the Brinkmann form, we notice that
for a line element of the form
\begin{equation}
 ds^2 = 2 \, du \, dv + \sum\limits_{i} A_i(u) \, dx^2_i \ ,
\end{equation}
we can  replace
\begin{equation}
\label{BrinkmannTransf}
 x_i \rightarrow \frac{x_i}{\sqrt{A_i}} \ , \qquad v \rightarrow v + \frac{1}{4} \sum\limits_{i} \frac{\dot{A}_i}{A_i} \, x^2_i \ .
\end{equation}
Implementing these, the line element takes the  form
\begin{equation}
 ds^2 = 2 \, du \, dv + \sum\limits_{i} dx^2_i +\Big( \sum\limits_{i} F_i(u) \, x^2_i \Big) \, du^2 \ ,
\end{equation}
where
\begin{equation}
 F_i = \frac{1}{4} \frac{\dot{A}^2_i}{A^2_i} + \frac{1}{2} \frac{d}{du} \Big(   \frac{\dot{A}_i}{A_i}  \Big) \ .
\end{equation}

In the particular case of (\ref{Metricpprxi}) we have
\begin{equation}
 A_z = \frac{\r^2}{\r^2 + 1} \ , \qquad A_w = \frac{\r^2}{\r^2 + 1} - 4 J^2 \ .
\end{equation}
Hence after replacing
\begin{equation}
  z \rightarrow \frac{z}{\sqrt{A_z}} \ , \qquad w \rightarrow \frac{w}{\sqrt{A_w}} \ , \qquad v \rightarrow v + \frac{1}{4} \, \frac{\dot{A}_z}{A_z} \, z^2 + \frac{1}{4} \, \frac{\dot{A}_w}{A_w} \, w^2 \ ,
\end{equation}
we get
\begin{equation}
\label{ds2ppNATD}
 \begin{aligned}
   ds^2 & = 2 \, du \, dv + d\bar{r}^2 + \bar{r}^2 \, d\Om^2_3 + dx^2 + x^2 \, d\b^2 +  dz^2 + dw^2
   \\[5pt]
   & - \Bigg[   \frac{\bar{r}^2}{16} + \frac{x^2}{16} \,  (8 J^2 - 1) + \frac{ (\r^2 + 1)^2}{\r^4} J^2 z^2 - F_z \, z^2 - F_w \, w^2  \Bigg] \, du^2 \ ,
 \end{aligned}
\end{equation}
where
%
\begin{equation}
  F_z = \frac{4 \, J^2 \big(   4 \, \r^2 + 1  \big) + 3 \, \big(   4 \, J^2 - 1  \big) \, \r^4}{4 \, \r^4 \, \big(   \r^2 + 1  \big)^2} \ ,
\qquad  F_w = - \frac{3}{4 \, \big(  \r^2 + 1  \big)^2} \ .
\end{equation}
The NS-NS and R-R fields become
\begin{equation}
\label{B2F4ppNATD}
 \begin{aligned}
  B_2 & = \frac{\r^2}{\sqrt{\big(  1 + \r^2   \big) \big(   1 - 4 J^2  \big) - 1}} \, dz \wedge dw + \frac{z}{ 2 \, \big(  \r^2 + 1  \big)} \, dw \wedge du
  \\[5pt]
  & + \frac{w \, \r^2}{ 2 \, \big(  \r^2 + 1  \big) \Big[\big(  \r^2 + 1   \big) \big(   1 - 4 J^2  \big) - 1\Big]} \, du \wedge dz \ ,
  \\[5pt]
  F_4 & = \frac{2 \, J \, x \, \sqrt{\r^2 + 1}}{\tilde{\tilde g}_s} \, du \wedge dx \wedge dz \wedge d\b \ .
 \end{aligned}
\end{equation}
The dilaton is given by \eqref{DilatonppNATD}.
The NS-NS field strength has a quite simple expression, 
\begin{equation}
 H = dB_2 = \frac{1}{2} \, \frac{\r^2 + 3}{\r^2 + 1} \, du \wedge dz \wedge dw \; .
\end{equation}
In general, one can consider different gauge choices for the NS two-form. 
In the analysis of the quantization of the string we make the choice
\be
B_2=- \frac{1}{2} \, \frac{\r^2 + 3}{\r^2 + 1} \Big(  \kappa_1 \, z \, du\wedge dw -\kappa_2 \, w \, du\wedge dz \Big),
\label{zaxa}
\ee
with $\kappa_1+\kappa_2=1$.

Next, we study the string quantization 
in the pp-wave of (\ref{ds2ppNATD})-(\ref{zaxa}).

\subsection{Closed string quantization on the pp wave}

Using the same notation for coordinates as in (\ref{coordCorresp}), the closed string action in the  background  of (\ref{ds2ppNATD})-(\ref{zaxa}) is
\begin{equation}
 \begin{aligned}
  S = & - \frac{1}{4 \, \pi \, \a'} \int d\tau \, d\s \Bigg[   \partial X^i \cdot \partial X^i +\Bigg(   \frac{\big(  X^1  \big)^2 + \big(  X^2  \big)^2 + \big(  X^3  \big)^2 + \big(  X^4  \big)^2}{16} 
  \\[5pt]
  & + \frac{\big(  X^5  \big)^2 + \big(  X^6  \big)^2}{16} \,  (8 J^2 - 1) + \frac{ (\r^2 + 1)^2}{\r^4} J^2 \big(  X^7  \big)^2 - F_z \, \big(  X^7  \big)^2 - F_w \, \big(  X^8  \big)^2  \Bigg)
  \\[5pt]
  & - \frac{\r^2 + 3}{\r^2 + 1} \, \Big(  \k_1 \, X^7 \, \partial_\s X^8 - \k_2 \, X^8 \, \partial_\s X^7   \Big)  \Bigg] \ .
 \end{aligned}
\end{equation}
Again we fixed the Weyl symmetry like in \eqref{WeylFix} and the residual diffeomorphism invariance by considering \eqref{DiffeoFix}.

The equations of motion for the scalars $X^i ,\, i = 1 , \ldots , 8$ now are
\begin{equation}
 \begin{aligned}
  & \Box X^i - \frac{1}{16} \, X^i = 0 \ , \qquad i = 1, \ldots , 4 \ ,
  \\[5pt]
  & \Box X^i - \frac{8 \, J^2 - 1}{16} \, X^i = 0 \ , \qquad i = 5, 6 \ ,
  \\[5pt]
  & \Box X^7 - \Bigg[  \frac{ (\r^2 + 1)^2}{\r^4} J^2 - F_z  \Bigg] \, X^7 + \frac{ 1 }{2} \, \frac{\r^2 + 3}{\r^2 + 1} \, \partial_\s X^8 = 0 \ ,
  \\[5pt]
  & \Box X^8 + F_w \, X^8 - \frac{ 1 }{2} \, \frac{\r^2 + 3}{\r^2 + 1} \, \partial_\s X^7 = 0 \ ,
 \end{aligned}
\end{equation}
where $\Box \equiv \eta^{\a \b} \partial_\a \partial_\b = - \partial^2_\tau + \partial^2_\s$. The corresponding boundary conditions are
\begin{equation}
 \begin{aligned}
  & \big( \partial_\s X^i \big) \, \d X^i \Big|_{\s = 0}^{\s = 2 \pi} = 0 \ , \qquad i = 1 , \ldots , 6 \ ,
  \\[5pt]
  & \Big(   \partial_\s X^7 +  \frac{ \kappa_2 }{2} \, \frac{\r^2 + 3}{\r^2 + 1} \, X^8  \Big) \, \d X^7 \Big|_{\s = 0}^{\s = 2 \pi} = 0 \ ,
  \\[5pt]
  & \Big(   \partial_\s X^8 - \frac{ \kappa_1 }{2} \, \frac{\r^2 + 3}{\r^2 + 1} \, X^7  \Big) \, \d X^8 \Big|_{\s = 0}^{\s = 2 \pi} = 0 \ .
 \end{aligned}
\end{equation}
Notice that generically, the masses (or frequencies) of the system depend on $\rho$. Hence, masses do depend on the light-cone time.
We will discuss this phenomenon in more generality in Section \ref{sectionQFT} and in Appendix \ref{appendixGM}.

From the formulas above it is obvious that the equations of motion for the first six scalars can be solved easily. 
We will not present the solutions for the scalars here, just the oscillator frequencies for each scalar, which we will need for the study of dual field theory. 
They are
\begin{equation}
 \begin{aligned}
  & \om^2_{n, i} = n^2 + \frac{1}{16} \ , \qquad i = 1 , \ldots , 4 \ ,
  \\[5pt]
  & \om^2_{n, i} = n^2 + \frac{8 \, J^2 - 1}{16} \ , \qquad i = 5, 6 \ .
 \end{aligned}
\end{equation}
There is no unstable (tachyonic) mode in the $(5^{th}, 6^{th})$ 
directions only if $J^2\geqslant 1/8$, as advertised.

\subsubsection{The limit $u\ll 1$}
We can solve explicitly the equations  in the limit $u\ll 1$, which however does not imply that $\rho\ll 1$. It is simply 
the starting region of the trajectory in the affine 
parameter of the wave. 
As an aside, notice that the opposite limit, 
$u\gg 1$ (that corresponds to $\rho\rightarrow \infty$), 
the solution becomes 
the Abelian T-dual solution 
treated previously (except for the dilaton, 
which is now non-constant, whereas the dilaton in the Abelian case is constant).
This correspondence was observed in \cite{Macpherson:2015tka}, 
\cite{Lozano:2016kum}.

In the $u\ll 1$ limit, $\rho$ becomes a constant, 
\begin{equation}
 \r_c = \frac{\sqrt{\big(  1 - 4 \, J^2  \big)^2 \, c^2_\r + 16 \, J^2}}{ 2 \, \sqrt{  1 - 4 \, J^2  }} \ .
\end{equation}
This can be further simplified by an appropriate choice of the integration constant $c_\r$, for example
\begin{equation}
  \r_c = \sqrt{\frac{1 + 4 \, J^2}{1 - 4 \, J^2}} \qquad \textrm{if} \qquad c_\r = \pm \frac{2}{1 - 4 \, J^2} \ .
\end{equation}

The pp-wave solution in the small $u$ limit becomes
\begin{equation}
 \begin{aligned}
  ds^2 & = 2 \, du \, dv + d\bar{r}^2 + \bar{r}^2 \, d\Om^2_3 + dx^2 + x^2 \, d\b^2 +  dz^2 + dw^2
   \\[5pt]
   & - \frac{1}{16} \, \Big[   \bar{r}^2 + x^2 \,  (1 - 2 \, a) + \big(4 - a^2\big) \, z^2 + 3 \, a^2 \, w^2  \Big] \, du^2 \ ,
   \\[5pt]
   e^{- 2 \, \Phi} & = \frac{\r^2_c + 1}{\tilde{\tilde g}_s^2} = \frac{2}{a \, \tilde{\tilde g}_s^2} \ ,
   \\[5pt]
   B_2 & = - \frac{a + 1}{2} \, \big(   \kappa_1 \, z \, du \wedge dw - \kappa_2 \, w \, du \wedge dz   \big) \ , \qquad \kappa_1 + \kappa_2 = 1 \ ,
   \\[10pt]
   F_4 & = \frac{2 \, \sqrt{2} \, J \, x}{\sqrt{a} \, \tilde{\tilde g}_s} \, du \wedge dx \wedge dz \wedge d\b \ ,
 \end{aligned}
\label{llxx}\end{equation}
with $a \equiv 1 - 4 \, J^2$. Notice that background in (\ref{llxx}) is an exact solution (on its own) of the type-IIA supergravity 
equations of motion.
In a similar way, the equations of motion for the scalars simplify to
\begin{equation}
 \begin{aligned}
  & \Box X^i - \frac{1}{16} \, X^i = 0 \ , \qquad i = 1, \ldots , 4 \ ,
  \\[5pt]
  & \Box X^i - \frac{1 - 2 \, a}{16} \, X^i = 0 \ , \qquad i = 5, 6 \ ,
  \\[5pt]
  & \Box X^7 - \frac{4 - a^2}{16} \, X^7 + \frac{ 1 }{2} \, \big(  1 + a  \big) \, \partial_\s X^8 = 0 \ ,
  \\[5pt]
  & \Box X^8 - \frac{3 \, a^2}{16} \, \, X^8 - \frac{ 1 }{2} \, \big(  1 + a  \big) \, \partial_\s X^7 = 0 \ .
 \end{aligned}
\end{equation}
The oscillator frequencies of the scalars are
\begin{equation}
 \begin{aligned}
  & \om^2_{n, i} = n^2 + \frac{1}{16} \ , \qquad i = 1 , \ldots , 4 \ ,
  \\[5pt]
  & \om^2_{n, i} = n^2 + \frac{1 - 2 \, a}{16} \ , \qquad i = 5, 6 \ ,
  \\[5pt]
  & \om^2_{n, \pm} = n^2 + \frac{1}{16} \, \big(   a^2 + 2  \big) \pm \frac{1}{8} \, \big(  a + 1  \big) \sqrt{16 \, n^2 + \big(  a - 1  \big)^2} \ .
 \end{aligned}
\label{lllxx}\end{equation}
This completes our analysis of the quantization of the string in the pp-wave
 corresponding to the 
non-Abelian T-dual of $AdS_5\times S^5$. We will now analyze the amount of supersymmetry preserved by the pp-wave background discussed in this section.

\section{Supersymmetry of the pp waves}\label{section3xx}

In this section we count the number of possible supersymmetries preserved by the pp-wave corresponding to the non-Abelian T-dual solution. 
For convenience we work with the pp-wave in the Brinkmann form and we define the coordinates $y_i , \, i = 1, \ldots , 8$ as
\begin{equation}
 d \bar{r}^2 + \bar{r}^2 \, d\Om^2_3 = \sum\limits_{i = 1}^4 (dy^i)^2 \ , \qquad dx^2 + x^2 \, d\b^2 = (dy^5)^2 + (dy^6)^2 \ , \qquad y^7 = z \ , \qquad y^8 = w \ .
\end{equation}
In these coordinates the non-Abelian T-dual pp-wave solution reads
\begin{equation}
 \begin{aligned}
   ds^2_{pp} & = 2 \, du \, dv + \sum\limits_{i = 1}^8 dy^2_i
    + \cH \, du^2 \ ,
    \\[5pt]
    H & = dB_2 =  \frac{1}{2} \, \frac{\r^2 + 3}{\r^2 + 1} \, du \wedge dy^7 \wedge dy^8 \ ,
    \\[5pt]
       F_4 & = - \frac{2 \, J \, \sqrt{\r^2 + 1}}{\tilde{\tilde g}_s} \, du \wedge dy^5 \wedge dy^6 \wedge dy^7 \ .
 \end{aligned}
\end{equation}
where now
\begin{equation}
\label{cH}
 \cH =  - \Bigg[   \frac{\vec{y}_4^2}{16} + \frac{(y^5)^2+(y^6)^2}{16} \,  (8 J^2 - 1) + \Bigg( \frac{(\r^2 + 1)^2}{\r^4} J^2 
 - F_z \Bigg) \, (y^7)^2 - F_w \, (y^8)^2  \Bigg] \ .
\end{equation}
Using the frame proposed in \eqref{ppWaveFrame} 
the formulas above take the form
\begin{equation}
 \begin{aligned}
   ds^2_{pp} & = 2 \, e^{+} e^{-} + \sum\limits_{i = 1}^{8} (e^i)^2 = \eta_{a b} \, e^a \, e^b \ ,
    \\[5pt]
    H & = dB_2 =  \frac{1}{2} \, \frac{\r^2 + 3}{\r^2 + 1} \, e^{-} \wedge e^7 \wedge e^8 \ .
    \\[5pt]
       F_4 & = - \frac{2 \, J \, \sqrt{\r^2 + 1}}{\tilde{\tilde g}_s} \, e^{-} \wedge e^5 \wedge e^6 \wedge e^7 \ .
 \end{aligned}
\end{equation}
Notice that this solution preserves a global $SO(4) \times SO(2)$ symmetry, where the $SO(4)$ factor corresponds 
to rotations in the $y^1 , \ldots , y^4$ directions and the $SO(2)$ factor to rotations in the $y^5, \, y^6$ directions.

For the susy analysis below we follow the conventions in appendix D of \cite{Itsios:2013wd}. 
Thus, the susy variations of the dilatino and gravitino in our case are
\begin{equation}
\label{SuSyIIA}
 \begin{aligned}
  \delta \l & = \frac{1}{2} \slashed{d}\Phi \,\e - \frac{1}{24} \slashed{H} \, \s_3 \, \e + \frac{e^{\Phi}}{8 \times 24} \, \slashed{F}_4 \, \s_1 \, \e \ ,
  \\[5pt]
  \delta \psi_{\mu} & = D_{\mu} \e - \frac{1}{8} \, H_{\mu \nu \r} \, \G^{\nu \r} \, \s_3 \, \e + \frac{e^{\Phi}}{8 \times 24} \, \slashed{F}_4 \, \s_1 \, \G_{\mu} \, \e \ ,
 \end{aligned}
\end{equation}
where
\begin{equation}
 D_{\mu} \e = \partial_{\mu} \e + \frac{1}{4} \, \om^{\,\,\,\, a b}_{\mu} \, \G_{a b} \, \e
\end{equation}
and the Killing spinor $\e$ consists of two Majorana-Weyl spinors $\e_\pm$, such that
\begin{equation}
 \e = \mqty(\e_+ \\ \e_-) \ .
\end{equation}
In the type IIA supergravity, the Killing spinor $\e$ satisfies the chirality condition 
$\G_{11} \e = - \s_3 \, \e$. We denote the Pauli matrices by $\s_i , \, i = 1, 2, 3$.

Due to the form of the spin connection, the covariant derivatives can be written as
\begin{equation}
 D_{+} = \partial_{+} \ , \qquad D_{-} = \partial_{-} + \frac{1}{4} \, \partial_i \cH \, \G_{+ i} \ , \qquad D_i = \partial_i \ ,
\end{equation}
where we assume the summation for repeated indices. Also, the equation \eqref{pmFrame} suggests that we should define the $\G^{\pm}$ matrices as
\footnote{
With this definition it is guaranteed that \[   \{   \G^0 , \, \G^0  \} = - \{   \G^9 , \, \G^9  \} = - 2 \, \mathbbm{1} \ , \qquad \{   \G^0 , \, \G^9  \} = 0 \ .  \]
}
\begin{equation}
 \G^{\pm} = \frac{\G^9 \pm \G^0}{\sqrt{2}} \ .
\end{equation}
From this formula we easily verify that $\big( \G^{\pm} \big)^2 = 0$.


\subsection{Susy of the pp-wave solution for motion along $\r$ and $\xi$}

For convenience we consider a more general form of the pp-wave where now the function $\cH$ is given by the expression 
(instead of the one in \eqref{cH})
\begin{equation}
\label{Hgeneral}
 \cH = \sum\limits_{i , j = 1}^8 F_{ij}(u) \, y^i \, y^j \ , \qquad F_{ij} = F_{ji} \ .
\end{equation}
Moreover, we turn on a NS three-form and a RR four-form with expressions
\begin{equation}
\label{H3F4general}
 H = 4 \, g(u) \, e^{-} \wedge e^7 \wedge e^8 \ , \qquad F_4 = f(u) \, e^{-} \wedge e^5 \wedge e^6 \wedge e^7
\end{equation}
and we also consider a non-trivial dilaton which is a function of $u$ 
only, i.e. $\Phi = \Phi(u)$. In the analysis that follows, 
we take the functions $F_{ij} , \, f , \, g$ and $\Phi$ to 
be arbitrary. If we want to make contact with the 
non-Abelian T-dual pp-wave of the previous section we just
restrict to the following expressions for those functions:
\begin{equation}
\label{susyansatz}
 \begin{aligned}
   & \Phi = - \frac{1}{2} \, \ln \Big[  \frac{\r^2 + 1}{\tilde{\tilde g}_s^2}  \Big] \ , \qquad f = - 2 \, J \, e^{- \Phi} \ , \qquad g = \frac{1}{8} \, \frac{\r^2 + 3 }{\r^2 + 1} \ ,
   \\[5pt]
   & F_{11} = F_{22} = F_{33} = F_{44} = - \frac{1}{16} \ , \qquad F_{55} = F_{66} = \frac{1 - 8 \, J^2}{16} \ ,
   \\[5pt]
   & F_{77} = F_z - \frac{(\r^2 + 1)^2}{\r^4} \, J^2 \ , \qquad F_{88} = F_w \ , \qquad F_{ij} = 0 \ , \quad \textrm{for} \quad i \ne j \ .
 \end{aligned}
\end{equation}
We now start our analysis with the dilatino equation.

\vskip 10pt

\noindent \textbf{The dilatino equation}

\vskip 10pt

If we plug the above ansatz into the dilatino variation \eqref{SuSyIIA} and require that it vanishes, we get
\begin{equation}
\label{DilatinoCond}
 \G^- \, \Big[  \dot{\Phi} - 2 \, g \, \G^{78} \, \s_3 + \frac{f \, e^{\Phi}}{4} \, \G^{567} \, \s_1   \Big] \, \e = 0 \ .
\end{equation}
For the functions in \eqref{susyansatz}, which correspond to the pp-wave of our interest, we realize that the only projection that guarantees the vanishing of the dilatino equation is $\G^- \e = 0$. This tells us that the pp-wave that we studied above preserves $16$ supercharges. A supersymmetry enhancement would imply the existence of more projections in the dilatino equation which should be also compatible with the gravitino variations.

For completeness, let us now turn to the variation of gravitino and examine it component by component (the analysis is done using the frame components and not the natural ones).

\vskip 10pt

\noindent \textbf{The $\delta \psi_{+}$ component}

\vskip 10pt

Since the NS three-form does not have legs on $\e^+$ and also since $\slashed{F}_4 \, \G_{+} = \slashed{F}_4 \, \G^{-} = 0$ (recall that $\slashed{F}_4$ already includes $\G^-$), the gravitino variation $\delta \psi_+$ reduces to
\begin{equation}
 \delta \psi_+ = D_+ \e = \partial_+ \e \ .
\end{equation}
Thus if we impose the vanishing of $\delta \psi_+$ we get that the Killing spinor $\e$ is independent of $v$, i.e. $\e = \e(u , y^i)$.

\vskip 10pt

\noindent \textbf{The $\delta \psi_{i} , \, i = 1 , \ldots , 8$ components}

\vskip 10pt

The vanishing of $\delta \psi_i$ implies
\begin{equation}
 \partial_i \e = \G^- \, \Big[    g \, \big(   \delta_{i8} \, \G^7 - \delta_{i7} \, \G^8  \big) \, \s_3 - \frac{f \, e^{\Phi}}{8} \, \G^{567} \, \G^i \, \s_1     \Big] \, \e \ .
\end{equation}
Notice that since $\partial_i \e$ is proportional to $\G^-$, if we act with one more derivative $\partial_j \ , \,\, j = 1 , \ldots , 8$ we obtain
\begin{equation}
 \partial_i \partial_j \e = 0 \ , \qquad i , j = 1 , \ldots , 8 \ .
\end{equation}
This tells us that the Killing spinor $\e$  is linear in $y^i$ and thus its solution is
\begin{equation}
 \e = \Bigg[   1 + \G^- \, \Big[    g \, \big(   \delta_{i8} \, \G^7 - \delta_{i7} \, \G^8  \big) \, \s_3 - \frac{f \, e^{\Phi}}{8} \, \G^{567} \, \G^i \, \s_1     \Big] \, y^i   \Bigg] \, \chi(u) \ ,
\end{equation}
where we assume the summation of the index $i$.

\vskip 10pt

\noindent \textbf{The $\delta \psi_{-}$ component}

\vskip 10pt

We now turn to the variation $\delta \psi_-$. Notice that in this case the covariant derivative $D_-$ becomes
\begin{equation}
 D_- = \partial_- + \frac{1}{2} \, F_{ij} \, y^j \, \G^{- i} \ .
\end{equation}
If we expand $\delta \psi_-$ then we find
\begin{equation}
 \begin{aligned}
  \delta\psi_- & = D_- \e - g \, \G^{78} \, \s_3 \, \e - \frac{f \, e^{\Phi}}{8} \, \G^{-} \, \G^{+} \, \G^{567} \, \s_1 \, \e
  \\[5pt]
  & = \partial_u \e + \frac{1}{2} \, F_{ij} \, y^j \, \G^- \, \G^i \, \e - g \, \G^{78} \, \s_3 \, \e - \frac{f \, e^{\Phi}}{8} \, \G^{-} \, \G^{+} \, \G^{567} \, \s_1 \, \e \ ,
 \end{aligned}
\end{equation}
where we took into account that
\begin{equation}
 \partial_{-} = \partial_u - \frac{1}{2} \, F_{ij} \, y^i \, y^j \, \partial_v
\end{equation}
and the fact that $\e$ does not depend on $v$. If we now require the vanishing of $\delta \psi_-$ and substitute the solution for $\e$ we find
\begin{equation}
\label{deltapsim}
 \begin{aligned}
  & \Bigg[   1 + \G^- \, \Big[    g \, \big(   \delta_{i8} \, \G^7 - \delta_{i7} \, \G^8  \big) \, \s_3 - \frac{f \, e^{\Phi}}{8} \, 
  \G^{567} \, \G^i \, \s_1     \Big] \, y^i   \Bigg] \, \partial_u \chi
  \\[5pt]
  & +  \G^- \, \Big[    \dot{g} \, \big(   \delta_{i8} \, \G^7 - \delta_{i7} \, \G^8  \big) \, \s_3 - \frac{\partial_u 
  ( f \, e^{\Phi}) }{8} \, \G^{567} \, \G^i \, \s_1     \Big] \, y^i \, \chi + \frac{1}{2} \, F_{ij} \, y^j \, \G^- \, \G^i \, \chi
  \\[5pt]
  &  - g \, \G^{78} \, \Bigg[   1 + \G^- \, \Big[    g \, \big(   \delta_{i8} \, \G^7 - \delta_{i7} \, \G^8  \big) \, 
  \s_3 - \frac{f \, e^{\Phi}}{8} \, \G^{567} \, \G^i \, \s_1     \Big] \, y^i   \Bigg] \, \s_3 \, \chi
  \\[5pt]
  & - \frac{f \, e^{\Phi}}{8} \, \G^{-} \, \G^{+} \, \G^{567} \, \Bigg[   1 + \G^- \, \Big[    g \, \big(   \delta_{i8} \, 
  \G^7 - \delta_{i7} \, \G^8  \big) \, \s_3 - \frac{f \, e^{\Phi}}{8} \, \G^{567} \, \G^i \, \s_1     \Big] \, y^i   \Bigg] \, \s_1 \, \chi= 0 \ .
 \end{aligned}
\end{equation}
From the last equation we obtain nine total conditions. In order to find those conditions we first isolate the piece that 
is independent of the $y^i$'s. This piece gives
\begin{equation}
\label{eqchi}
  \partial_u \chi - g \, \G^{78} \, \s_3 \, \chi - \frac{f \, e^{\Phi}}{8} \, \G^{-} \, \G^{+} \, \G^{567} \, \s_1 \, \chi = 0 \ .
\end{equation}
The other eight conditions come from the requirement that the coefficient of each of the $y^i$'s in \eqref{deltapsim} must vanish. 
Hence, for each $y^i$, after eliminating $\partial_u \chi$ using \eqref{eqchi} we get
\begin{equation}
 \begin{aligned}
  & \G^- \, \Big[    \dot{g} \, \big(   \delta_{i8} \, \G^7 - \delta_{i7} \, \G^8  \big) \, \s_3 - \frac{\partial_u ( f \, e^{\Phi}) }{8} \, \G^{567} \, 
  \G^i \, \s_1 + \frac{1}{2} \, F_{ij} \, \G^j     \Big] \, \chi
  \\[5pt]
  & + \G^- \, \Big[ \frac{f \, e^{\Phi}}{4} \, \G^{567} \, (i \s_2) - 2 \, g \, \G^{78} \Big] \, \Big[    g \, \big(   \delta_{i8} \, \G^7 - \delta_{i7} \, 
  \G^8  \big) \, \s_3 - \frac{f \, e^{\Phi}}{8} \, \G^{567} \, \G^i \, \s_1     \Big]\, \s_3 \, \chi = 0 \ .
 \end{aligned}
\end{equation}

This completes our analysis of the preserved supersymmetry. 
Both Abelian and non-Abelian plane waves preserve sixteen supercharges.
In the following section, we will discuss different aspects of the field theory dual to the backgrounds in Sections \ref{section1xx} and \ref{section2xx}.

\section{Quantum Field Theory interpretation}\label{sectionQFT}

Here we present a field theoretical 
interpretation for the results of the strings 
on the pp waves studied in previous sections. 
We will mostly refer to the Abelian pp-waves in Section 
\ref{section1xx}, but we will also discuss the non-Abelian 
case of Section \ref{section2xx} as well.

\subsection{The AdS/CFT map}
We start by discussing the field theories dual to the backgrounds 
in (\ref{maka1})
and (\ref{NATDsolNS})-(\ref{NATDsolRR}).

The field theory dual to the Abelian and non-Abelian backgrounds 
was considered in 
\cite{Lozano:2016kum}. 
The field theory dual to the background in (\ref{maka1})---the
T-dual of $AdS_5\times S^5/\mathbb{Z}_k$-- 
is a circular quiver with $k$ nodes and gauge group $SU(N)^k$. 
The theory has ${\cal N}=2$ supersymmetry in 3+1 dimensions.  
At each node we have a ${\cal N}=2$ vector multiplet and 
a ${\cal N}=2$ bifundamental hypermultiplet between each two adjacent nodes.
 
In the non-Abelian case, the field theory dual proposed 
in  \cite{Lozano:2016kum} is an infinitely-long quiver, 
with increasing gauge group, i.e. 
$SU(N)\times SU(2N)\times SU(3N)\times ...\times SU(kN)\times ...$. 
The quiver terminates only when a 'completion' of the 
background is considered, at an arbitrary position 
$\tilde{\rho}=P \pi $ 
by the addition of a suitable flavour group.  
After this completion  each node has an ${\cal N}=2$ 
vector multiplet, and each link between 
two adjacent nodes contains an ${\cal N}=2$ 
bifundamental hypermultiplet (hence satisfying the condition 
for conformality $N_f=2N_c$ for each node).
Similar completions have been found in different 
backgrounds obtained using non-Abelian T-duality, 
see for example the papers in \cite{completions}.

In the context of Penrose limits and pp-waves, the  
quiver field theory dual to $AdS_5\times S^5/\mathbb{Z}_k$ 
has been considered  in the papers \cite{Alishahiha:2002ev} and 
\cite{Mukhi:2002ck}. 
As we will see discuss, our Penrose limits in Sections
\ref{section1xx} and \ref{section2xx} are different. 
Yet, they still have common ground 
with the analyses in  \cite{Alishahiha:2002ev} and 
\cite{Mukhi:2002ck}. 
The T-duality will play an important role. 
A T-dual picture was also considered 
in \cite{Mukhi:2002ck}, nevertheless the order of 'operations' 
is important. Indeed,  the paper \cite{Mukhi:2002ck} 
considered first a Penrose limit, and then T-duality, 
whereas in this work we consider T-duality first, and then a Penrose 
limit. Another important difference is the scaling of 
the parameters, which in our work is different to that in the papers
\cite{Alishahiha:2002ev} and
\cite{Mukhi:2002ck}.

{\bf Field theory limit}

We have seen that in both the Abelian and non-Abelian cases, in order for the rescaled coordinates $\psi$ and $\rho $ in 
(\ref{rescc})  and in (\ref{rhosmall}) to be of order one, we must have that 
 \be
k\sim \frac{L^2}{\a'}=\sqrt{4\pi g_s^BN}\;,
\ee
which means that
\be
\frac{g^2_{YM}N}{k}=\frac{4\pi g_s^B N}{k}\sim k \rightarrow \infty.
\ee
Here $g_s^B$ is the string coupling of the type IIB theory 
(before the T-duality), related to the string coupling 
in the type IIA theory by 
\be
g_s^A=g_s^B\frac{\sqrt{\a'}}{L}\Rightarrow g_s^B=g_s^A\frac{L}{\sqrt{\a'}}.
\ee
Note that in the Abelian case, the effective string coupling 
is $\tilde g_s$. Using (\ref{gsredefx}), we have
\be
\frac{k^2}{N}\sim g_s^B= 
g_s^A\frac{L}{\sqrt{\a'}}=\tilde g_s \frac{L^2}{\a'}\sim \tilde g_s k\;.
\ee
This implies
\be
\tilde g_s \sim \frac{k}{N}.
\ee
If we keep fixed the YM coupling, it means $g_s^B$ is fixed, 
leading to $N\sim k^2$, and then
\be
\tilde g_s\sim \frac{1}{k}.
\ee

In the non-Abelian case, the effective string coupling 
$\tilde{\tilde g}_s$ was defined in (\ref{gsredefxx}). 
In this case we have
\be
\frac{k^2}{N}\sim g_s^B= g_s^A\frac{L}{\sqrt{\a'}}
=\tilde{\tilde g}_s \frac{L^4}{\a'^2}= 4\pi \, \tilde {\tilde g}_s \, g_s^BN\;,
\ee
which in the limit of fixed $g_s^B$ leads to
\be
\tilde{\tilde g}_s\sim \frac{1}{N}.
\ee

 Both results imply that strings on the pp waves are classical, 
i.e. they cannot break or join. 
The only relevant computation is the calculation 
of eigenergies of the strings, to all orders in $\a'$. 

As discussed in Section \ref{section3xx}, the Abelian and non-Abelian
pp waves have 
only the standard 1/2 supersymmetry, there is no susy enhancement. 
We should expect the same from the field theory analysis. 

{\bf \underline{Various Penrose limits of $AdS_5\times S^5/\mathbb{Z}_k$ 
and their field theory dual}}

Below, we will compare our pp wave with the cases previously 
considered in the bibliography \cite{Alishahiha:2002ev}, \cite{Mukhi:2002ck}
and \cite{Kuperstein:2003jd}. 

In the paper \cite{Alishahiha:2002ev} it was pointed out that 
we can have two possible 
Penrose limits for $AdS_5\times S^5/\mathbb{Z}_k$:
\begin{itemize}
\item{Along a direction different from that 
of the $\mathbb{Z}_k$ identification, leading to an 
${\cal N}=2$ supersymmetric result. 
Basically, this is 
the $\mathbb{Z}_k$ orbifold of the pp wave of maximal supersymmetry 
(the Penrose limit of the $AdS_5\times S^5$ background). The field theory 
was analyzed in \cite{Alishahiha:2002ev}. 
In this case, one has the field theory scaling
\be
\frac{g^2_{YM}N}{k}=\frac{4\pi g_s^BN}{k}\rightarrow \infty\;,
\ee
and one considers also a  large R-charge limit for operators, 
but the scaling does not 
affect the amount of supersymmetry of the sector, which is still 
${\cal N}=2$.}

\item{Along the direction of the $\mathbb{Z}_k$  identification, 
leading to a maximally supersymmetric background. 
The result of the Penrose limit has  a 
periodic $x^-$ ("lightcone space") direction. 
This was analyzed by \cite{Mukhi:2002ck}. In this case one 
considers the scaling
\be
\frac{g^2_{YM}N}{k}=\frac{4\pi g_s^BN}{k}={\rm fixed}\;,
\ee
together with a large R-charge limit on operators.
However, now the large R-charge limit, coupled with the fixed ratio above, leads to a supersymmetry 
enhancement to maximal for the field theory sector. }
\end{itemize}

Our study of Sections \ref{section1xx} and \ref{section2xx}
corresponds to a Penrose limit along the direction 
of the  $\mathbb{Z}_k$  identification, 
yet with the scaling limit and amount of 
supersymmetry of the first item above. 
This means that we will have a different field theory sector 
from the one analyzed in \cite{Alishahiha:2002ev}, \cite{Mukhi:2002ck}
and 
\cite{Kuperstein:2003jd}. 

{\bf \underline{The field theory and its symmetries}}

In order to understand the field theory sector dual to our pp-waves, 
we first find useful to
summarize the properties of the full quiver field theory. 
We focus on the scalars, since they are the easiest to 
identify with the gravitational side. 

The scalar fields are: an 
adjoint complex scalar $X_i$ at each node $i$ (belonging to 
the vector multiplet) 
and the two bifundamental complex scalars
$V_i$ and $W_i$, between nodes $i$ and $i+1$, one of them 
in the $(i,\overline{i+1})$, the other in the complex conjugate 
$(\overline i,i+1)$ representation of the $i$-th gauge group.
The superpotential, in ${\cal N}=1$ superfields notation, is
\be
W=\sum_{i=1}^k\int d^2\theta \, \Tr_{i+1}[V_i X_iW_i]\;,
\ee
where the trace is necessary, because the term is in the gauge 
group of the $i+1$ node (the $i$ node indices are all contracted).
The kinetic terms are
\be
L_{kin}= \sum_{i=1}^k \int d^2\theta \, d^2\bar \theta \, 
\Tr_i [\bar V_i e^{-2V}V_i+W_ie^{+2V}\bar W_i + X_i^\dagger e^{V} X_i].
\ee

The field theory symmetries are:
\begin{itemize}
\item{ An $SU(2)_R$ that rotates $V_i$ and $\overline W_i$ 
(the two complex scalars in the hypermultiplet
that transform in the same representation of the gauge groups).
It also rotates the fermionic $d^2\theta$-terms. 
Under this symmetry, the fields $(V_i,\bar W_i)$ and $(\bar V_i,W_i)$ 
are doublets in conjugate representations.
Inside this $SU(2)_R$, there is a $U(1)$-symmetry with Cartan generator $J_1$ 
(denoted by $J'$ in \cite{Mukhi:2002ck})}.

\item{An $U(1)_R$ that rotates only the chiral field 
$X_i$ inside the ${\cal N}=2$ vector multiplet 
and $d^2\theta$-terms, leaving the bifundamentals invariant, 
\be
X_i\rightarrow e^{i\a} X_i\;,\;\;\;\;
d^2\theta\rightarrow e^{-i\a}d^2\theta.
\ee}

\item{An extra (non-R) $U(1)$ that multiplies 
the bifundamental scalars, $V_i,W_i$, by the opposite phases $e^{i\a}$,
\be\label{nonr}
V_i\rightarrow e^{i\a }V_i\;,\;\;\;\;
W_i\rightarrow e^{-i\a}W_i.
\ee}
\end{itemize}
the generator of which is denoted by $J_2$ (this was called $J$ 
in \cite{Mukhi:2002ck}).  

Having sumarized the field content and symmetries of the 
quiver field theory, let us discuss now how these are realized 
in our backgrounds and the corresponding pp-waves.

{\bf \underline{The gravity dual and identification of 
coordinates and symmetries}}

Our pp waves
correspond to taking first a T-duality
along a direction made periodic by a $\mathbb{Z}_k$ 
identification, and then taking the 
Penrose limit along the same direction. In the non-Abelian case
we dualize on an $SU(2)$-isometry containing the $\mathbb{Z}_k$-identified 
coordinate. 

The procedure that \cite{Mukhi:2002ck} followed
is in reverse order. Indeed,  they took first the Penrose limit 
in the $\mathbb{Z}_k$ 
direction, and then considered the Abelian 
T-dual interpretation (for T-duality on the same $\mathbb{Z}_k$-identified 
direction). Let us analyze the
Abelian case in what follows.

From the form of the metric in our  case, 
the T-dual of $AdS_5\times S^5/\mathbb{Z}_k$---see (\ref{maka1}), 
we see that the background has explicit 
symmetries: $SU(2)\times U(1)$, where the $SU(2)$ is the symmetry of  the 
$S^2$ parametrized by the angles ($\chi$,$\xi$), and the $U(1)$ corresponds 
to translations in the angle $\b$. 
There is also a shift symmetry corresponding to 
 translations along the compact coordinate $\psi$, 
which takes values in $[0,2\pi k]$, 
and has a periodicity $2\pi k$. This corresponds to another $U(1)$ 
symmetry\footnote{Notice that we should consider the 
field strenth $H_3=dB_2$
for the symmetry to be apparent.}. Therefore, in total, 
we have the isometries 
$SU(2)_{\chi,\xi}\times U(1)_\b \times U(1)_\psi$, which are identified with 
the symmetries of the field theory. Indeed,  
the identification of the dual gravity coordinates 
with field theory scalars can be achieved by embedding 
the five-sphere of (\ref{ads5xs5}) into Euclidean 
space. In order to do this, we parametrize the six dimensional space
in terms of three complex scalars 
$Z_{i}, i=1,2,3,$ given explicitly by
\be
Z_1=L \, \sin \a \, e^{i\beta} \ , \quad
Z_2=L \, \cos \a \cos\chi \, e^{i(\xi+\psi)} \ , \quad
Z_3=L \, \cos \a \sin \chi \, e^{i(\xi-\psi)} \ .\label{zangles}
\ee

Here $\sum_i |Z_i|^2=L^2$, $L$ corresponds to the size of the space, 
 and the coordinate $\psi$ has periodicity $2\pi /k$. 
Indeed, $Z_1$ is the scalar inside the vector multiplet, that we denote by 
$X$, and $Z_2,Z_3$ are the bifundamentals, denoted by $V,W$ above.
 

We will now identify symmetries  of our quiver 
field theory:
\begin{itemize}

\item{The $SU(2)_R$ acts on the scalars $Z_2,Z_3$. Indeed, it acts 
on the $S^2$ angles described by $( \chi,\xi)$
in our parametrization of the gravity dual (after T-duality). 
 This means that in particular, our gravity dual symmetry 
$U(1)_{\xi}\subset SU(2)_R$ gives a charge $J_1$. }

\item{The $U(1)_R$ acts on the scalar $Z_1$ 
by multiplication by a phase $e^{i\a}$, and $Z_1$ is 
identified with our $X$-superfield, which means the 
field theory symmetry $U(1)_R$ corresponds to the $U(1)_\beta$ in (\ref{zangles}). }

\item{ The extra $U(1)$ acts on the scalars 
$Z_2,Z_3$ (corresponding to our $V,W$) 
by $Z_2\rightarrow e^{i\a}Z_2$, $Z_3\rightarrow e^{-i\a}Z_3$, which means 
that $\psi\rightarrow \psi+\alpha,$ 
in (\ref{zangles}). This is just a shift in the direction 
of orbifolding, $\psi$.
Therefore this extra symmetry, $U(1)_\psi$, corresponds to the 
non-R $U(1)$ the action of which was defined in (\ref{nonr}).}
\end{itemize}
The charge assignments are summarized in  Table \ref{charges}.
We see that the resulting value of the energy defined by \cite{Mukhi:2002ck},
$H=2p^-=\Delta-kJ-J'$ (which plays the role of $"\Delta-J"$ in the BMN limit) 
is $H=0$ for $V_i$ and 
$H=1$ for $W_i$ and $X_i$. This 
allows us to consider $V_i$ as the ground state and $W_i$, $X_i$ as 
the oscillator modes.

\begin{center}
\label{charges}
  \begin{tabular}{ | l || c | r |  r |   r |  r |   r |}
    \hline
     & $X$ & $V$&$W$& $\overline{X}$&$\overline{V}$&$\overline{W}$  \\ \hline
    $\Delta$ & 1 & 1&1&1&1&1 \\ \hline
    $J_1$ & 0 & $\frac{1}{2}$&$\frac{1}{2}$&0&$-\frac{1}{2}$&$-\frac{1}{2}$ \\ \hline 
    $k J_2$ & 0 & $\frac{1}{2}$&$-\frac{1}{2}$&0&$-\frac{1}{2}$&$\frac{1}{2}$ \\ \hline 
           $H$ & 1 & 0&1&1&2&1 \\ \hline 
  \end{tabular}
  \captionof{table}{quantum numbers (charges) of the bosonic operators and their conjugates}
\end{center}
This finishes our analysis of symmetries.
Let us discuss the states and operators in the field theory dual 
to the type IIA string on the Abelian pp-wave.

\subsection{Abelian case: operators and field theory sector}

As described in the previous subsection, we expect the 
field theory operators corresponding to our pp wave to be the same ones 
considered by  \cite{Mukhi:2002ck}-\cite{Kuperstein:2003jd}, 
since the Penrose limit is taken in the same direction. 
However, the 
scaling of parameters and amount of supersymmetry is different, 
so we expect the eigenenergies and eigenstates to be different. 

The operators we want to describe 
have a large charge in the direction corresponding to
the gravity dual symmetry $U(1)_\psi$, which 
is $U(1)_{\rm extra}$ in the field theory. 
The scalars charged under it are $V_i$ and $\bar W_i$ (its  
charge is denoted by $J$ in the work 
\cite{Mukhi:2002ck}). We also have a large charge in the direction 
corresponding to the gravity dual symmetry
$U(1)_\xi\subset SU(2)_R$ (called $J'$ in \cite{Mukhi:2002ck}). 
The ratio of the two charges, $J'/J$ in \cite{Mukhi:2002ck}, 
is identified with---see (\ref{papax})-(\ref{papaxx}),
\be
\left(\frac{J'}{J}\right)_{\rm them}=\left(\frac{J_1}{J_2}\right)_{\rm us}=
\frac{\dot \xi}{\dot \psi}=\left.\frac{2J}{\sqrt{1-4J^2}}\right|_{\rm us}\;.
\ee
We have used the values for the null geodesic around which 
we expand. The minimum value for the quantity is reached when our parameter 
$J$ 
takes its minimum value, $J=1/(2\sqrt{2})$, 
when we get $(J_1/J_2)=1$, and the maximum 
is reached when ou $J$ takes its maximum value, 
$J=1/2$, when we get $(J_1/J_2)=\infty$. 

From the list of charges in  Table \ref{charges},
we see that there is no way to vary $J_1/J_2$, since 
we need the vacuum to have zero energy $H=2p^-=0$, and the 
only field with that property is $V_i$, which has $J_1=J_2=1/2$, 
corresponding for us
to $J_{}=1/(2\sqrt{2})$. It is not clear how to see 
this restriction of the value of $J_{}$ from our pp wave analysis. 

As explained also in \cite{Mukhi:2002ck}, 
we construct the vacuum using only $V_i$-fields, 
that must wrap completely around the circular
quiver. The reason is that the $V_i$'s 
are bifundamental between the adjacent gauge groups 
$SU(N)_i\times SU(N)_{i+1}$. Such a vacuum has a 
"winding" around the quiver. 
The authors of  \cite{Mukhi:2002ck} associated the vacuum with 
a state of momentum $p=1$ and winding $m=0$. 
However, because of the T-duality between our picture and theirs, their state
\be
a^\dagger_n |p,m\rangle
\ee
where $p$ is momentum and $m$ is winding along the $\mathbb{Z}_k$ (lightcone) direction, and $n$ is a transverse direction, becomes
\be
a^\dagger_n |m,p\rangle
\ee
in our T-dual case, thus exchanging momentum with winding. 

That means that the state of zero-winding 
and one-lightcone unit of momentum in 
\cite{Mukhi:2002ck},  
becomes a state of winding-one and zero-lightcone momentum for 
us, corresponding to the operator winding once around the quiver. 
More formally, we write
\be
|p=1,m=0\rangle_{\rm them}=|m=0,p=1\rangle_{\rm us}={\cal O}_k=\frac{1}{\sqrt{\cal N}}\Tr[V_1V_2...V_k]\;,
\ee
with conformal dimension $\Delta =k=J_1+J_2$ ($\Delta=J+J'$ in the Type IIB
analysis of \cite{Mukhi:2002ck}), 
so with zero energy, $H=0$. 
A state of winding $p$ (momentum $p$ in Type IIB) just winds $p$ times around 
the quiver. 

To insert oscillators on these states with winding, we look for the fields with energy $H=2p^-=1$, which are:

- as usual, the 4 covariant derivatives $D_a$, $a=0,1,2,3$. 

- the bifundamentals $W_i , \, \bar W_i$

- the vectors $X_i, \, \bar X_i$. 

These are the usual 4+4 bosonic oscillators also present in the standard BMN case. Note however that
$D_a$, $X_i$ and $\bar X_i$ can be inserted  inside the operators 
(at the site $i$), whereas an insertion of
$W_i$ needs also an extra insertion of $V_i$, 
in order to return  to the same gauge group $SU(N)_i$ 
(both $W_i$ and $V_i$ are bifundamental,
but with opposite 'orientations'). 
Similarly, an insertion of $\bar W_i$ necessitates the removal 
of the corresponding $V_i$ inside the operator,
since  $\bar W_i$  acts in the same way as $V_i$. 
Then the insertions of 
$D_a,X_i$ and $\bar X_i$ at zero transverse momentum 
give
\bea
{\cal O}_{D_p}=a^\dagger_{D,0}|m=0,p=1\rangle_{\rm us}&=&
\frac{1}{\sqrt{Nk}}\frac{1}{\sqrt{\cal N}}
\sum_{i=1}^k\Tr[V_1...V_{i-1}(D_aV_i)...V_k] \ ,
\nonumber\\[5pt]
{\cal O}_{{X}_p}=
a^\dagger_{X,0}|m=0,p=1\rangle_{\rm us}&=&\frac{1}{\sqrt{Nk}}\frac{1}{\sqrt{\cal N}}\sum_{i=1}^k\Tr[V_1...V_{i-1}X_iV_i...V_k] \ ,
\\[5pt]
{\cal O}_{\bar{X}_p}=
a^\dagger_{\bar X,0}|m=0,p=1\rangle_{\rm us}&=&\frac{1}{\sqrt{Nk}}
\frac{1}{\sqrt{\cal N}}\sum_{i=1}^k\Tr[V_1...V_{i-1}\bar X_iV_i...V_k] \ ,
\nonumber
\label{xxzz}\eea
whereas the insertions of $W_i$ and $\bar W_i$ give
\begin{equation}
 \begin{aligned}
   {\cal O}_{W,0}=
a^\dagger_{W,0}|m=0,p=1\rangle_{\rm us}&=\frac{1}{\sqrt{N^2k}}\frac{1}{\sqrt{\cal N}}\sum_{i=1}^k\Tr[V_1...V_{i-1}V_iW_iV_i...V_k] \ ,
\\[5pt]
{\cal O}_{\bar{W},0}=
a^\dagger_{\bar W,0}|m=0,p=1\rangle_{\rm us}&=\frac{1}{\sqrt{k}}
\frac{1}{\sqrt{\cal N}}\sum_{i=1}^k\Tr[V_1...V_{i-1}\bar W_iV_{i+1}...V_k] \ .
\label{opexx}
 \end{aligned}
\end{equation}
To insert a field of mode number $n$, we put phases as usual, for instance
\be
a^\dagger_{X,n}|m,p=1\rangle_{\rm us}=\frac{1}{\sqrt{Nk}}\frac{1}{\sqrt{\cal N}}\sum_{l=1}^k\Tr[V_1...V_{l-1}X_lV_l...V_k]e^{\frac{2\pi il n}{k}}.
\ee
The important difference is that now putting 
several $n_i$, the sum of all gives our momentum 
(the winding in the Type IIB analysis), 
\be
m=\sum_i n_i.
\ee

There are no states of vanishing momentum in the Type IIB analysis.
This is because there is no vanishing momentum 
state in Discrete Light Cone Quantization, DLCQ.
As a consequence, there are no operators of  vanishing winding 
in the Type IIA description. 
Indeed, we cannot construct a state of zero energy with charges 
$J_1=J_2\neq 0$ other than from $V_i$'s. We could imagine using $(V_i\bar V_i)$ pairs, but those have vanishing total charge and dimension 2, 
so $H=2p^-=2$, changing the energy of the state. 

Let us understand better 
the absence of zero-winding operators in the Type IIA pp-wave.

{\bf \underline{Windings on the pp wave}}

Here, we provide an interpretation, from the point of view of the 
string theory on the pp wave, of the fact that 
we {\em need} winding in our states. Specially
those states associated with the (T-duality) $\psi$-direction, 
and no winding in the other directions. Below we present 
a plausability argument explaining this fact.

As is well known, strings in flat space with some compact directions
and with a constant Neveu-Scharz B-field along those directions
change their spectrum (and its states have non-zero
winding).
Indeed, in the book \cite{Polchinski:1998rq} 
(more precisely in Volume I, around page 250), the effect
is studied. We suggest that a  similar mechanism may be at work here,
even when we are not working in flat space and our B-field is not constant.
For this, we need to show that some of the directions of our pp-wave in (\ref{abppxx1})-(\ref{abppxx2})
can be considered periodic.

To do this, 
consider the boundary conditions in (\ref{bdconds}).
When we have {\em closed strings},  
for the transverse coordinates $X^7, X^8$ ($z$ and $w$), 
we cannot put $\delta X^7=0$  
and $\delta X^8=0$ at $\sigma=0$ and $2\pi$, which would be 
an open string boundary condition. 
Of course, we can put $X^7(\sigma=2\pi)=X^7(\sigma=0)$ 
{\em and} $X^8(\sigma=2\pi)=X^8(\sigma=0)$, but 
more precisely, we must first impose 
$\delta X^7(\sigma=2\pi)=\delta X^7(\sigma=0)$ 
and $\delta X^8(\sigma=2\pi)=\delta X^8(\sigma=0)$. 

Then we obtain the boundary conditions
\begin{equation}
 \begin{aligned}
  & \Big(\d_\sigma X^7+\frac{\mu \tilde \kappa_2}{2}X^8\Big)\Big|_{ \sigma=0}^{\sigma=2\pi}=0 \ ,
\\[5pt]
  & \Big(\d_\sigma X^8-\frac{\mu \tilde \kappa_1}{2}X^7\Big)\Big|_{\sigma=0}^{\sigma=2\pi}=0.
 \end{aligned}
\label{fafaxx}
\end{equation}

If we work in the gauge where the coefficient $\kappa_2=0$, 
we can take the solutions to these equations such that
\be
X^7(\sigma=2\pi)=X^7(\sigma=0)+ C,
\ee
which implies that the first of (\ref{fafaxx}) is satisfied. 
The second equation would indicate the non-periodic character of $X^8$.

One can see the winding of the lightcone direction by
%
deriving the periodicities 
on the pp wave. From the first relation in 
(\ref{consts}), $t\propto u$, we see that lightcone 
time $u$ is not periodic. Note that in our case, we have (choosing $8J^2=1$),
\begin{equation}
 \begin{aligned}
  & c_1=\frac{1}{4} \ , \qquad
c_2=J=\frac{1}{2\sqrt{2}} \ , \qquad
c_3=\sqrt{1-4J^2}=\frac{1}{\sqrt{2}} \ ,
\\[5pt]
& c_4=\frac{\sqrt{1-4J^2}}{2}=\frac{1}{2\sqrt{2}} \ , \qquad
c_5=-2J=-\frac{1}{\sqrt{2}} \ , \qquad
c_6=\frac{2}{\sqrt{1-4J^2}}=2\sqrt{2} \ .
 \end{aligned}
\end{equation}
From this periodicity and the periodicity of $\psi$ by 
$2\pi \a' k/L^2$, we obtain 
\begin{equation}
 \begin{aligned}
   v & \sim v+\frac{2\pi \a' }{L^2}\frac{L^2 k}{c_6}=v+\frac{2\pi \a' k}{2\sqrt{2}}
   \\[5pt]
 \end{aligned}
\end{equation}
That means that the lightcone space direction $v$ 
is compactified with radius $\a' k$. 

In summary, we have made plausible the argument that the presence 
of the B-field on a compact direction adds winding to the states.
As we mentioned, it also changes the energy levels. To this we move now.



{\bf \underline{Eigenenergies}}

In (\ref{freqabx}), we discussed  that at $n=0$, 
the frequencies of the closed string moving on the pp-wave of the 
Abelian background are
\be
\omega_{0,a}=\frac{1}{4}\;, \;\; a=1,2,3,4\;,\;\;\;
\omega_{0,i}=\frac{\sqrt{8J^2-1}}{4}\;,\;\;\; i=5,6\;,\;\;
\omega_{0,+}=J ,\;\;
\omega_{0,-}=0.
\label{freqdis}
\ee
We should obtain the same 
results from the energies $H$ of the operator states described above, 
at $n=0$. 

The insertion of $D_a$ on the state of zero-momentum and winding-one 
gives an energy $H=1$, since $D_a$ has no charges, 
but has mass dimension $\Delta=1$. 
This corresponds to the $\omega_{0,a}$. 
Naively the insertions of $X, \bar X, W,\bar W$ would also give 
energies $H=1$, and this does not seem to match. 

An important difference with the analysis of \cite{Mukhi:2002ck}, 
where $g^2_{YM}N/k$ was kept finite, is that for us this quantity
 is divergent.

This leads to an absence of 
the cancellations that allow for a simple result. 
In our case, there will be interactions that cannot be neglected, 
that will correct the eigenenergies. 
We did not study the calculations of the Feynman diagrams, 
since they are very involved.
We will just present below a sketchy argument, showing
 the type of interactions we expect to 
be relevant to change the spectrum. 

Because of the ${\cal W}\sim g_{YM}\Tr_{i+1}[V_i X_i W_i]$ 
term in the superpotential, 
we have the interaction  term in the scalar potential
\be
V\sim g^2_{YM}\Tr_i|W_i V_i|^2=\Tr_i (\bar W_i  W_i V_i \bar V_i)\;,
\label{epdmxx}\ee
and this term will mix ${\cal O}_{W,0}$ and ${\cal O}_{\bar W,0}$, 
the operators with zero momentum 
insertions of $W$ and $\bar W$ defined in (\ref{opexx}). 
Indeed, introduce in ${\cal O}_{\bar W,0}$ 
the "particle-antiparticle" pair $V_i \bar V_i$
before $\bar W_i$, which would annihilate, 
though they have a perturbative "energy" of $\Delta -J_1-J_2=2$, so 
this would seem to change the energy of the state. Since 
the effective coupling $g^2_{YM} N/k$ is large, 
interaction effects will 
be dominant, and they can plausibly 
reduce the energy of the state with "particle-antiparticle" pair insertion. 
Then the insertions of $(W_i V_i)$ and $(\bar V_i \bar W_i)$ 
could mix via the interaction term from the potential described
above.

That means that the energy levels, which start off equal 
at $g^2_{YM}=0$, could split off into a higher one and a lower one, 
as observed in the 
pp wave---see (\ref{freqdis}). 

Similarly, the same superpotential term 
$g_{YM}\Tr_{i+1}[V_i X_i W_i]$ leads also to the potential term
\be
V\sim g^2_{YM}\Tr_k[\bar X_i \bar V_i V_i X_i]\;,
\ee
which will mix ${\cal O}_{X,0}$ and ${\cal O}_{\bar X,0}$ in (\ref{xxzz}), 
the last with the introduction of a particle-antiparticle pair $\bar V_i V_i$. 
That means that the eigenenergies of these states could also be corrected.

\subsection{Non-Abelian case: RG flow}

In the case of the pp wave obtained from the non-Abelian T-dual solution, 
as we have already observed, the pp wave at large $\rho$, obtained
from the $u\gg 1$ limit ($u$ being the 
lightcone time on the pp wave), reduces to the 
Abelian pp wave, except for the varying dilaton---compare 
(\ref{abppxx1})-(\ref{abppxx2}) 
with (\ref{ds2ppNATD}) and (\ref{DilatonppNATD}).
However, as was also pointed out in \cite{vanGorsel:2017goj}, the matching becomes 
exact if we instead consider $\rho=\rho_0+\psi$ and take the scaling limit
\be
\rho_0\rightarrow \infty, \quad \tilde{\tilde g}_s\rightarrow \infty,\quad  \frac{\rho_0}{ \tilde{\tilde g}_s}=\textrm{fixed}\equiv \frac{1}{\tilde{g_{s}}}.
\label{coupxx}
\ee
On the other hand, at small $u$ ($u\ll 1$), 
leading to a finite $\rho=\rho_c$, the pp wave solution is given 
in (\ref{llxx}), and the frequencies of the 
string modes were calculated in (\ref{lllxx}). 

This means that the pp wave solution, 
and the corresponding eigenenergies for closed string modes 
on the pp wave, {\it "flow"}
in the lightcone time $u$, from $u=0$ to $u=\infty$. This flow not only appears in the Penrose limit of the Sfetsos-Thompson solution. It
is indeed a more generic phenomenon for certain solutions of the Gaiotto-Maldacena type \cite{Gaiotto:2009gz}, as we explain in Appendix \ref{appendixGM}. Also, 
in Appendix \ref{janusappendix}, we see a similar flow in the Penrose limit of the Janus solution, dual to a defect CFT. 

The idea of an RG flow related to a nontrivial pp wave 
was considered before, in \cite{Corrado:2002wi} and \cite{Gimon:2002sf}. 
However, those authors considered simply 
an RG flow in the radial coordinate  
of the gravity dual. They discussed  Penrose limits 
near geodesics moving on different 
coordinates, both in the IR and in 
the UV of the holographic RG flow. 

Here, we have a different situation: 
we have an RG flow for the masses
(or frequencies) of string modes, $\omega=\omega(u)$, between the 
UV and the IR.
This is obtained from the way in which we took 
the Penrose limit. We are actually realizing the
picture advocated (but not obtained) in  
\cite{Corrado:2002wi} and \cite{Gimon:2002sf}. 

The flow is  between a theory in the UV, 
which corresponds to $\rho\rightarrow\infty$ or $u\rightarrow \infty$, and the IR theory, at $u\rightarrow 0$. 
Indeed, we see that at large $\rho$, the equivalent 
Abelian coupling $\tilde g_s\rightarrow 0$ (corresponding to the UV), 
by the identification in (\ref{coupxx}) above with the non-Abelian case. 
This UV fixed point theory 
is conformal, namely it is the Abelian quiver field theory 
(which is ${\cal N}=2$ superconformal). 
The IR fixed point theory is defined by the $u\ll 1$ limit. 

We can naturally ask: how is it possible to have an RG flow, 
when the gravity dual has an AdS factor, dual to conformal symmetry? 
We can provide two possible non-exclusive answers:
\begin{itemize}
\item{ One answer is that we must 
now consider {\em the full string theory, including winding modes}, 
since we consider the T-dual to $AdS_5\times S^5/\mathbb{Z}_k$, and 
winding modes correspond to momentum modes before T-duality. In the 
full string theory, the conformal invariance is broken by these 
winding modes, which make possible the flow. This is not 
fully satisfactory, as there
are winding modes in the Abelian T-dual plane wave, 
and no flow in frequencies. Nevertheless being both backgrounds and
dynamics different, the winding modes on the non-Abelian plane wave
might have the effect argued above.}
\item{Another possibility is that the non-Abelian 
T-dual background in (\ref{NATDsolNS})-(\ref{NATDsolRR}) 
is actually dual to a non-conformal
field theory in higher dimensions, reducing itself to a four dimensional CFT
only for large and small values of the $\rho$-coordinate. We explore this 
deconstruction possibility in the next sub-section. This picture is also supported by the material in Appendix \ref{janusappendix}.}
\end{itemize}

Note that, since in our Penrose limit we 
identify $t=u$, and AdS space is 
in global coordinates, meaning that on the boundary we have 
$S^3\times \mathbb{R}_t$. Then $t=u$ is the radial time in the field 
theory, and $e^{\pm t}=e^{\pm u}$ is the radius in the Euclidean field theory, 
i.e. distance scale
(or energy scale). Thus the flow in $u$ is really 
an RG flow, that is flow in energy (or distance) scale. 

Unfortunately, there isn't much we can calculate in the field theory 
corresponding to the non-Abelian case; we must consider the string theory calculations
as simply {\em predictions} for the field theory. 

Notice that the non-Abelian T-dual background does not
have the $U(1)_{\mathrm{extra}}$. This symmetry only appears
in the metric for large values of the $\rho$. 
We expect the operators corresponding to the string states 
to be  extensions to the new (infinite) quiver of the 
operators in the Abelian case. 
Then the excitations are extensions of those in the Abelian case.
Correspondingly we find the same string 
eigenenergies {\em in the UV} (at large $u$). 
But in the non-Abelian case, we have a more interesting dynamics, 
leading to the RG flow in eigenenergies. 

We now discuss the second alternative bullet-point explanation of the 
flow in the non-Abelian plane wave.
\subsection{Deconstruction}

The large quiver describing the field theory dual to the non-Abelian T-dual background
is reminiscent of the "dimensional deconstruction" ideas 
of \cite{ArkaniHamed:2001ca} and 
\cite{ArkaniHamed:2001nc}. 

In "deconstruction",
at intermediate energies  a fifth dimension appears. More precisely,
after a particular choice of Higgsed  vacuum, a KK tower
of states associated with a compact dimension appears. 
As usual in KK theories, at even lower energies 
(smaller than the inverse  radius of the 
compact dimension), the fifth dimension becomes invisible again. 
The UV theory is the full 4-dimensional quiver theory as the 
effects of the VEV become negligible. 

In the previous subsection we discussed that
there is an RG flow in $u$ between the UV theory (the quiver) 
and the IR theory. The deconstruction picture suggests
that for intermediate $u$, $u\sim 1$, we should see a higher dimensional
field theory. 

We have seen that the quiver implied long spin chain operators describing a discretized dimension. The normal interpretation for the spin chain is as a 
discretized string worldsheet in the gravity dual, but in the field theory itself it can now be also described as an extra dimension, since as we saw we 
need to consider only operators with "winding" around it. 

We now make more precise in which sense the background
in (\ref{NATDsolNS})-(\ref{NATDsolRR}) 
represents a higher dimensional field theory.


\subsubsection{Wilson loops and signs of extra dimensions}
To give  support to the proposal that the non-Abelian of $AdS_5\times S^5$, the Sfetsos-Thompson solution, is actually representing a 
field theory in dimension higher than four, we will calculate the potential between two static charges. We shall observe that, 
in a certain regime, there are deviations from the
usual $V_{QQ}\sim 1/L_{QQ}$ characteristic of four dimensional CFTs. 

We first present the intuitive argument. The solution in (\ref{NATDsolNS})-(\ref{NATDsolRR}) was interpreted in \cite{Lozano:2016kum}
as an {\it strictly infinite} length quiver $SU(N)\times SU(2N)\times SU(3N)\times...$. This CFT has infinite central charge, 
hence it is not a four dimensional field theory.
 The treatment of \cite{Lozano:2016kum} 
was to {\it complete} the quiver to make sense of it as a four dimensional CFT.
 This leads us to consider that the solution in  (\ref{NATDsolNS})-(\ref{NATDsolRR}) might
be dual to a field theory in higher dimension, realising ideas similar to those in deconstruction.

To make this more precise, we will consider the Sfetsos-Thompson solution with the $AdS_5$ space written in 
Poincare patch, with coordinates $[x_{1,3}, R]$. We shall  calculate the Wilson loop for a heavy quark-antiquark 
pair placed on the space directions $[x,\rho]$, separated by distances $L_x,L_\rho$, respectively. This is inspired 
by the fact that $\rho$ plays the role of the field space in the interpretation of  \cite{Lozano:2016kum}. We shall then 
consider a configuration
\bea
& & t=\tau, \;\;\; R=\sigma,\\
& & x=x(\sigma),\;\;\; \rho=\rho(\sigma).\nonumber
\label{configxxx}
\eea
In this configuration, the distance between the pair of quarks is $L=\sqrt{L_x^2+L_\rho^2}$. In the following, for 
computational purposes, we restrict ourselves to the 
regime in which the distance $L_\rho$ is much smaller than $L_x$, hence $L\sim L_x$. The action of the 
Nambu-Goto string is (we set $L=\alpha'=1$ and choose to place the string at  the value of the coordinate $\alpha=0$)
\bea
S_{NG}= \int dR \sqrt{1+ R^4 x'^2 + R^2 \rho'^2}.\label{actixxx}
\eea
We use the conserved quantities
\bea
& & J=\frac{\partial {\cal L} }{\partial \rho'}= \frac{R^2 \rho'}{ \sqrt{1+ R^4 x'^2 + R^2 \rho'^2}}.\nonumber\\[5pt]
& & M=\frac{\partial {\cal L} }{\partial x'}= \frac{R^4 x'}{ \sqrt{1+ R^4 x'^2 + R^2 \rho'^2}}.\label{conservedxx}
\eea
We will now produce two 'string effective actions' . One will be useful for small values of separations $L_{QQ}\sim 0$, the second will be trustable for strings that explore large parts of the bulk.

{\it \underline{Short strings, Large $R$ effective action}}

Using (\ref{conservedxx}), we replace $\rho' (J, R, x')$ in the action to obtain the action for an 'effective' string propagating on the non-Abelian T-dual background of (\ref{NATDsolNS})-(\ref{NATDsolRR}),
\bea
S= \int dR \sqrt{\frac{R^2}{R^2-J^2} + \frac{R^6}{R^2-J^2}x'^2 }=\int dx \sqrt{   \frac{R^6}{R^2-J^2}+ \frac{R^2}{R^2-J^2} R'^2   }.\label{wilsonxx}
\eea
This effective action represents
a string that moves in a background dual to 
a relevant deformation of ${\cal N}=4$ SYM. The {\it effective} background is singular
at $R=J$, hence long strings probing that position are not 
faithfully  described by this effective action.

For small $J$ and/or for large $R$, the action approximates 
that of the string on $AdS_5$. This is indeed intuitively correct, 
since $J$ measures how much the string explores the $\rho$-direction. 

We write now the expressions for the separation 
$L_{QQ}\sim  L_x$ (a good approximation when $\frac{L_\rho}{L_x}$ is small) 
and the Energy $E_{QQ}$ in terms of $R_0$, the point where the 
Nambu-Goto string turns back in the bulk,
\bea
& & L_{QQ}= \frac{2R_0^3}{\sqrt{R_0^2-J^2} }\int_{R_0}^{\infty} \frac{1}{R^2}\frac{dR}{\sqrt{   \frac{R^6}{R^2-J^2}-  \frac{R_0^6}{R_0^2-J^2}  }  },\\[5pt]
& & E_{QQ}=\int_{R_0}^\infty \frac{R}{\sqrt{R^2-J^2}}\left(\frac{R^3}{\sqrt{R^2-J^2}}\frac{1}{ \sqrt{   \frac{R^6}{R^2-J^2}-  \frac{R_0^6}{R_0^2-J^2} } } -1        \right)
-\int_{J}^{R_0} \frac{R}{\sqrt{R^2-J^2} }dR.\nonumber
\label{mmmvvvv}
\eea
Since we want to consider a string that extends 
very little in the $\rho$-direction, 
we expand the expressions in (\ref{mmmvvvv}), to first order in $J^2$.
The integrals can be exactly calculated. We obtain
\bea
& & L_{QQ}=\frac{1}{R_0}\left(l_{AdS} +l_1\frac{J^2}{R_0^2}  +.... \right),\\[5pt]
& & E_{QQ}=R_0\left(\epsilon_{AdS} + m_1\frac{J^2}{R_0^2} +....     \right).\nonumber
\label{EqqLQQ}
\eea
In this equation above, the numbers $l_{AdS}, \epsilon_{AdS}$ 
are those appearing in the original 
calculation \cite{Maldacena:1998im}. 
On the other hand, $l_1, \epsilon_1$ are just 
numerical constants that appear when evaluating the integral. 
The expansion in (\ref{EqqLQQ}) can 
be continued to higher orders. The same structure is preserved.

Inverting $R_0(L_{QQ})$ we can write
\bea
E_{QQ}\sim -\frac{C}{L_{QQ} (1-\frac{J^2}{l_{AdS}}  L_{QQ}^2)}.
\label{finalxx}
\eea
Notice that this expression is trustable for small values of 
$L_{QQ}$, effective strings that barely explore the bulk of the space. 
The point to be kept in mind is that 
there is a deviation from the CFT behaviour.

We will now analyze long strings that penetrate deeply into the bulk.

{\it \underline{Long strings, Small $R$ effective action}}

We consider (\ref{conservedxx}) and define
\bea
R^2\rho'^2= \frac{J^2}{M^2} R^6 x'^2,
\eea
and replace this in the action of (\ref{actixxx}). We obtain
\bea
S= \int dR \sqrt{1+ (R^4+\frac{J^2}{M^2} R^2) x'^2}=\int dx \sqrt{R'^2 +R^4(1+\frac{J^2}{M^2}R^2)}.
\eea
This effective action corresponds to a 
string moving in a background dual to ${\cal N}=4$ SYM 
deformed by an irrelevant operator 
(this is  reflected by the change in the large $R$ 
asymptotics of the metric). 

We should only trust these string configurations for large values 
of $L_{QQ}$.
In this case we write integrals for the separation and energy of the quarks pair, that are the analog expressions to those 
in (\ref{mmmvvvv}). We expand these integrals for small values of $x=J/M$ and obtain
\bea
& &L_{QQ}= \frac{1}{R_0}\left(l_{AdS} - l_1 x^2 R_0^2  \right),\nonumber\\[5pt]
& & E_{QQ}=R_0\left(\epsilon_{AdS} - x^2 R_0^2 \epsilon_1  \right).
\label{dddxx}
\eea
Here $l_1,\epsilon_1$ are numerical constants.
It is important to notice that the sub-leading terms in this expansion are not convergent (unless we cutoff the space), 
the expansions cease to make sense for smaller values of $L_{QQ}$
or larger values of the coordinate $R$. From (\ref{dddxx}) we obtain
\bea
E_{QQ}\sim- \frac{C_1}{L_{QQ}} -\frac{C_2}{L_{QQ}^3}.
\eea
Here $C_1,C_2$ are numerical constants.
This equation is again showing deviations from the characteristic 
conformal law.

In conclusion, we have made the point that the Wilson loop 
calculation shows indications that the Sfetsos-Thompson solution 
in (\ref{NATDsolNS})-(\ref{NATDsolRR}),
 is actually dual to a field theory in dimension higher than four. 
For either very short or very long strings, we recover a four 
dimensional conformal behaviour, but for strings that have a size that
explore the $\rho$-coordinate) we deviate from it.
 This suggest a form of deconstruction and provides some understanding to our flow within the pp-wave. In fact, this is the 
 reason why we have constant masses/frequencies for the plane wave in (\ref{ds2ppNATD}) both for large and small 
 values of the light-cone time $u$.
 In some sense, the result of  the 
non-Abelian T-dual plane wave suggests a defect-CFT behaviour. 
To compare, we calculate in the Appendix \ref{janusappendix}
the Penrose limit of the Janus solution, dual to a defect-CFT.
We will find similarities with the non-Abelian pp-wave.

\section{Conclusions}\label{conclu}
To close this paper, we present some final remarks.

We have found the plane wave geometries associated with the Abelian
and non-Abelian T-duals of $AdS_5 \times S^5$. These plane waves 
are consistent Type IIA backgrounds for a window of parameters 
and preserve sixteen supercharges.

We quantised the weakly coupled type IIA string on the Abelian T-dual
plane wave and wrote the equations describing the frequencies
of the non-Abelian T-dual case.
The unconventional feature is that the non-Abelian plane wave 
presents a 'flow' in the frequencies (betwen two constant values).

After a careful analysis of the isometries of the backgrounds
and global symmetries of the dual field theories we proposed a precise form 
for the BMN sector of operators described by the plane waves. Scaling limits
were analyzed. We gave intuitive 
arguments to explain the reason why our strings
generate states with non-zero winding. Using deconstruction ideas,
we presented arguments to explain the above-mentioned flow in 
eigenfrequencies for the non-Abelian T-dual plane wave.

It would be interesting to explore our BMN sector in more detail.
In particular, to have more control over the necessity of 
non-zero winding and the 
discrepancy in eigen-energies for the string on the plane wave. Indeed, 
it is important to understand precisely the corrections discussed around 
(\ref{epdmxx}). On the other hand, the proposal for deconstruction
and our calculation supporting it should be further explored  and extended.

It would be interesting to investigate the plane waves associated with
other backgrounds obtained via non-Abelian T-duality. 
In order to study similarities and differences we suggest to study 
the backgrounds found in \cite{Lozano:2012au},
\cite{Lozano:2013oma},  \cite{Lozano:2014ata},
\cite{Macpherson:2014eza}, \cite{Bea:2015fja}, 
\cite{Lozano:2015bra}, \cite{Lozano:2015cra}. 
These seem to be the natural 
candidates given their isometries.

This paper exploits a combination between non-Abelian T-duality
in the context of supergravity and the study of full the string
theory sigma model in (a limit of) that background. It should be interesting
to push forward ideas that exploit the synergy between non-Abelian 
T-duality and an all order sigma model calculation. This kind of 
studies might illuminate the nature of non-Abelian T-duality.

\section*{Acknowledgements}

We thank M. Blau for a useful correspondence.
The work of GI is supported by FAPESP grant 2016/08972-0 and 2014/18634-9. HN is supported in part by CNPq grant 304006/2016-5 and FAPESP 
grant 2014/18634-9. HN would also 
like to thank the ICTP-SAIFR for their support through FAPESP 
grant 2016/01343-7. HN would also like to thank the Swansea
Physics Department for hospitality during the period when this project was started, and to the Royal Society for Newton Mobility Grant 
NI160034 to visit Swansea. 
C.N is Wolfson Fellow of the Royal Society. S. Z. is a Newton International Fellow of the Royal Society.


\appendix

\section{Various ways to take the Penrose limit}\label{appendixA}

The Penrose limit corresponds to focusing near a null geodesic for motion in a curved spacetime, and Penrose's theorem says we should obtain a 
pp wave in the limit. 

{\bf Method 1}

In the original form, defined by Penrose, we follow the steps:

-Rescale all coordinates in the metric by the common scale $L$ of the metric, $x^\mu\rightarrow x^\mu/L$.

-Boost along a direction $\mu$. In lightcone coordinates $x^\pm =\mu\pm t$, this boost 
corresponds to $x^+\rightarrow e^\b x^+$ and $x^-\rightarrow e^\b x^-$. 
Together with the rescaling, we have $\mu+t\rightarrow e^b x^+/L$ and $\mu-t\rightarrow e^{-\b}x^-/L$.

-Identify the scale with the boost, $L=e^\b$, and take this quantity to infinity. We obtain $\mu+t=x^+$ and $\mu-t=x^-/L^2$, or (with an extra factor of 2)
\be
x^+=\frac{t+\mu}{2} \, , \quad x^-= L^2 \, \frac{\mu-t}{2} \, ,
\ee
which implies the expansion for the original coordinates
\be
t=x^+-\frac{x^-}{L^2} \, , \quad \mu=x^++\frac{x^-}{L^2} \, .
\ee

Then for the derivatives we obtain $\d_+=\d_t+\d_\mu$ and $\d_-=(\d_\mu-\d_t)/L^2$. In the context of the pp wave correspondence, 
one obtains the field theory map (here $p^-$ is conjugate to $x^+$ and $p^+$ to $x^-$)
\be
\frac{p^-}{\mu}=\Delta -J \, , \quad \mu p^+=\frac{\Delta+J}{L^2} \, .
\ee

{\bf Method 2}

But it was noted, originally by Tseytlin, that if $\mu$ is periodic (like it is often the case), the expansion for $t$ and $\mu$ above implies that the 
lightcone time $x^+=(t+\mu)/2$ is periodic, which is not good. Instead, one can write the expansion
\be
x^+=t \, , \quad x^-=L^2(\mu-t) \, ,
\ee
which implies
\be
t=x^+ \, , \quad \mu=x^++\frac{x^-}{L^2} \, .
\ee
In this form, lightcone time is identified with the usual time, which is not periodic, so we have a consistent story. 

Then also $\d_+=\d_t+\d_\mu$ and $\d_-=\d_\mu/L^2$, which leads to the field theory map
\be
\frac{p^-}{\mu}=\Delta-J \, , \quad
\mu p^+=\frac{J}{L^2} \, .
\ee

Both ways of taking the Penrose limit lead to the same pp wave metric in cases of interest. We can verify explicitly, that both in the case of $AdS_5
\times S^5$, and in the Janus case (next subsection), the resulting limits coincide. More generally, this is always true for the nontrivial terms, 
with $(du)^2$ times coordinates squared, since they come from the leading terms in the the expansion of $t$, $\mu$ (which are unchanged), 
times the subleading terms in the transverse coordinates (which are also unchanged).

In a more general case, we can always fix the transformation with $x^+=t$ to have the same Penrose limit as the usual one, since we have the 
same number of conditions as extra constants to be fixed. 

The only difference in this correct method amounts to the periodicities of the various coordinates, corresponding to periodicities in field theory. 
But these are important in our analysis, so in the text we will always use this method for taking the Penrose limit.

{\bf Method 3}

However, for completeness, we note that we could in principle also take the "maximally wrong" limit, and put all periodicity in $x^+$, and none 
in the lightlike space coordinate. This would mean 
\be
x^+=\mu \, , \quad x^-=L^2(\mu-t) \, ,
\ee
so also
\be
\mu=x^+ \, , \quad t=x^+-\frac{x^-}{L^2} \, .
\ee

In this case
$\d_+=\d_t+\d_\mu$ and $\d_-=-\d_t/L^2$, so we find for the field theory map
\be
\frac{p^-}{\mu}=\Delta-J\;,\;\;\;
\mu p^+=\frac{\Delta}{L^2}\;,
\ee
but the metric obtained from the limit is the same as for the first two methods, or in general can be put to the same form.

{\bf Generalization to motion in 2 coordinates}

Consider a Penrose limit in coordinates $(t,\xi,\rho)$ defined by (analogous to method 3 above)
\begin{equation}
 \begin{aligned}
dt & =\a  \Big(\frac{du}{2}+\b \, \frac{dw}{L} \Big)-\frac{dv}{L^2} \, ,
\\[5pt]
d\xi & = \delta \, du+\frac{dw}{L} \, ,
\\[5pt]
d\rho & =\gamma \, d u \, ,
 \end{aligned}
\end{equation}
where $u$ is lightcone time, and $v$ is lightcone space. 

The new (correct) Penrose limit, with no periodic time, is defined by
\begin{equation}
 \begin{aligned}
  dt & = \a\frac{du}{2} \, ,
  \\[5pt]
d\xi & = \delta du+\lambda \frac{dw}{L} \, ,
  \\[5pt]
d\rho & = \gamma du -\eta \frac{dw}{L}+\theta \frac{dv}{L^2} \, .
 \end{aligned}
\end{equation}

Since the leading terms ($L$-independent) in the Penrose expansion of $(t,\rho,\xi)$ are not corrected, being fixed by the need to cancel the ${\cal O}(L^2)$ 
terms in the metric, it means that the nontrivial Penrose terms, proportional to $du^2$ and with coordinates squared as factors, are the same. 

Therefore all we need is to have the same result for the terms
\be
L^2 \, \big[ -4dt^2+d\rho^2+d\xi^2 \big] \, .
\ee

First, the leading order, ${\cal O}(L^2)$, terms cancel in the same way as before, since as we said, that fixed the leading terms in $dt,d\xi,d\rho$:
\be
-\a^2+\gamma^2+\delta^2=0.
\ee
Next, the condition for the first subleading, yet still infinite, ${\cal O}(L)$ terms to cancel gives a condition on the coefficients. In the original 
transformation, this was 
\be
L \, du \, dw \, \big( -4 \, \a^2\b+2 \,\delta \big)=0.
\ee
Now, it is 
\be
2L \, du \, dw \, \big( \lambda \, \delta - \, \eta\,  \gamma \big)=0 \quad \Rightarrow \quad \frac{\eta}{\lambda}=2\frac{ \, \a^2\b}{\gamma}.
\ee

We can also impose the condition that the constant term proportional to $dw^2$ to have the same coefficient. Before, we had 
\be
dw^2 \, \big[ -4 \, \a^2 \, \b^2 + 1 \big] \, ,
\ee
whereas now we have
\be
dw^2 \, \big[ \lambda^2+\eta^2 \big] \, ,
\ee
so we obtain 
\be
\lambda^2+\eta^2=1-4 \, \a^2 \b^2.
\ee
But with the condition $\eta= 2 \, \a^2 \b \, \lambda/\gamma$ from before, we get
\begin{equation}
  \lambda =\gamma \sqrt{\frac{1-4\a^2\b^2}{\gamma^2+4\a^4\b^2}} \, ,
  \qquad
\eta = 2\frac{ \, \a^2 \, \b \, \lambda \,}{\gamma} .
\end{equation}

The condition that the constant term proportional to $dudv$ is the same is as follows. Before, the term was 
\be
4 \, \a \, du \, dv = 2 \, du \, dv \, ,
\ee
and now it is 
\be
2 \, \theta \, \gamma \, du \, dv \, ,
\ee
so we need 
\be
\theta =\frac{2 \, \a}{\gamma} \, .
\ee

As we see, in this case we had 3 equations for 3 unknowns, and the same matching will be true for motion in more general direction, involving several 
coordinates.

\section{More general CFTs: Gaiotto-Maldacena  backgrounds}\label{appendixGM}
One puzzling result of (\ref{ds2ppNATD}) is that the masses for the pp-wave obtained from the Sfetsos-Thompson 
background depend on the affine parameter $u$. This  motivates a question
about the genericity of such phenomenon. In other words: is this a particularity of the Sfetsos-Thompson solution or is it more general? 
To investigate this, we study here the same type of Penrose limit discussed above, but applied to backgrounds classified by 
Gaiotto and Maldacena \cite{Gaiotto:2009gz}. These solutions are dual to linear quiver field theories with ${\cal N}=2$ SUSY. 
The correspondence between CFT and geometry was very clearly spelled in \cite{ReidEdwards:2010qs}.
Below, we start with a brief summary of the formalism developed in studying these backgrounds and then 
discuss the Penrose limit and pp waves obtained from them.

The Gaiotto-Maldacena solutions  
can be written in terms of a potential function $V(\sigma,\eta)$. 
In particular, denoting $V'=\partial_{\eta} V$ 
and $\dot{V}=\sigma\partial_{\sigma}V$, 
one can write the Type IIA 
generic 
solution as
\bea
& & ds_{IIA,st}^2=\alpha'\left(\frac{2\dot{V} -\ddot {V}}{V''}\right)^{1/2}
\left[4 AdS_5 +\mu^2\frac{2V'' \dot{V}}{\Delta} 
{d \Omega^{2}_2(\chi,\xi)}\right.\cr
&&\left.\hspace{2cm}+\mu^2\frac{2V''}{\dot{V}}  
(d\sigma^2+d\eta^2)+ \mu^2\frac{4V'' \sigma^2}{2\dot{V}-\ddot{V}} 
d{\beta}^2 \right], \nonumber\\[5pt]
& & A_1=2\mu^4\sqrt{\alpha'}
\frac{2 \dot{V} \dot{V'}}{2\dot{V}-\ddot{V}}d{\beta},\;\;\;\; 
e^{4\Phi}= 4\frac{(2\dot{V}-\ddot{V})^3}{\mu^{4}V'' \dot{V}^2 \Delta^2}, 
\quad {\Delta = (2 \dot{V} - \ddot{V}) V'' + (\dot{V}')^2} \ ,  \nonumber \\[5pt]
& & B_2=2\mu^2\alpha' \left(\frac{\dot{V} \dot{V'}}{\Delta} -\eta\right) 
d\Omega_2,\;\;\; {A}_3={-} 4\mu^4 \alpha'^{3/2}
\frac{\dot{V}^2 V''}{\Delta}d{\beta} \wedge d\Omega_2.
\label{metrica}
\eea
The radius of the space is $\mu^2\alpha'=L^2$. The parameter $\mu$ is related to
the number of branes present in the solution.
In the following we  set $\mu=1$, with no important 
effect on the results. 
The  two-sphere $d \Omega^{2}_2(\chi,\xi)$ is parametrised by 
the angles $\chi$ and $\xi$ with corresponding volume 
form $d\Omega_{2}= \sin\chi d\chi \wedge d\xi$. The usual definition
$F_4= dC_3 + A_1\wedge H_3$ is also used.

The problem of writing  IIA/M-theory  solutions in this class, reduces
to  finding the function $V(\sigma,\eta)$ that solves
a Laplace equation with a given charge density $\lambda(\eta)$ and boundary conditions
\bea
& & \partial_\sigma[\sigma \partial_\sigma V]+\sigma \partial^2_\eta V=0,\;\;\;\;\;\;\;\lambda(\eta)= \sigma\partial_\sigma V(\sigma,\eta)|_{\sigma=0}\; .
\label{ecuagm1}\\[5pt]
& & \lambda(\eta=0)=\lambda(\eta=N_c)=0.\nonumber
\eea
Interestingly, the background and fluxes depend on $\dot{V}$,
$\dot{V}'$, ${\ddot{V}}$ and $V''=-\sigma^{-2}\ddot{V}$. 
Hence, given $\dot{V}$, we have all that is needed to write the 
Type IIA background.

The backgrounds discussed in Sections \ref{section1xx} and \ref{section2xx} turn out to be particular cases of this general class of solutions.
Indeed,  for the Abelian T-dual background in (\ref{maka1}),  
after redefining 
\be
\psi=r=\frac{2L^2}{\alpha'}\eta,\;\; \sigma=\sin\alpha,
\label{redefinition}
\ee 
the potential $V(\sigma,\eta)$ is
\be
V_{ATD}=\log\sigma-\frac{\sigma^2}{2} +\eta^2.
\label{potentialdensityTD}
\ee
For the non-Abelian T-dual background 
in \eqref{maka1}, redefining variables 
again as in (\ref{redefinition}), 
 the potential function  is
 \be
V_{NATD}=\eta\left(\log\sigma -\frac{\sigma^2}{2}\right)+
\frac{\eta^3}{3}.\label{potentialdensityNATD}
\ee 
In general, given a solution to the Laplace-like equation (\ref{ecuagm1}) that satisfies the boundary conditions, we have a background
dual to a well defined CFT. We will now study the pp-wave obtained from these backgrounds.

\subsection{Penrose limits and pp waves of Gaiotto-Maldacena backgrounds.}

It will be convenient in what follows to define the functions
\bea
& & f_1=\left(\frac{2\dot{V} -\ddot {V}}{V''}\right)^{1/2},\;\; f_2= \frac{2V'' \dot{V}}{\Delta} ,\;\;\; f_3= \frac{2V''}{\dot{V}} ,\;\;\;\nonumber\\[5pt]
& & f_4= \frac{4V'' }{2\dot{V}-\ddot{V}} ,\;\;\; f_5=2\left(\frac{\dot{V} \dot{V'}}{\Delta} -\eta\right) ,\;\;\; f_6
= \frac{2 \dot{V}\dot{V'}}{2\dot{V}-\ddot{V}},\;\;\; f_7=\frac{4\dot{V}^2V''}{\Delta},\nonumber\\[5pt]
& & {\Delta = (2 \dot{V} - \ddot{V}) V'' + (\dot{V}')^2}.\label{functkk}
\eea
In terms of these functions, we can write the background in (\ref{metrica}) as
\bea
& & \frac{ds^2}{L^2}= f_1(\sigma,\eta)\Big[ 4 AdS_5 + f_2(\sigma,\eta) d\Omega_2(\chi,\xi)+ f_3(\sigma,\eta)(d\sigma^2
+d\eta^2)+ f_4(\sigma,\eta)\sigma^2 d\beta^2\Big],\nonumber\\[5pt]
& & B_2= L^2 f_5(\sigma,\eta) \sin\chi d\chi \wedge d\xi, \;\;\; e^{4\Phi}= \frac{f_1^6 f_3^2}{ \Delta^2},\nonumber\\[5pt]
& & A_1=L f_6 d\beta,\;\;\;A_3=L^3 f_7 \sin\chi d\chi\wedge d\xi\wedge d\beta.
\label{mmmaaa}
\eea
We now study the geodesics of this background.
\subsection{Study of geodesics}
Following the procedure in previous Sections, we propose a possible geodesic that moves on
\be
\rho_{AdS}=\sigma=0,\;\;\; \chi=\frac{\pi}{2}.
\ee
The coordinates depend on the affine parameter of the geodesic $u$ according to
\be
t=t(u),\;\;\; \eta=\eta(u),\;\;\; \xi=\xi(u).
\ee
We obtain a Lagrangean
\be
L= f_1(0,\eta)\Big[ -4 \dot{t}^2 + f_3(0,\eta)\dot{\eta}^2 +f_2(0,\eta) \dot{\xi}^2 \Big].
\ee
After using the two conserved quantities $E, J$, we find the equation for $\eta(u)$ is
\bea
& & \dot{t}=  \frac{E}{f_1(0,\eta)},\;\;\; \dot{\xi}=\frac{J}{f_1(0,\eta) f_2(0,\eta)},\nonumber\\[5pt]
& & \dot{\eta}^2=\frac{1}{f_1(0,\eta)^2 f_3(0,\eta)} \left(\frac{E^2}{4}-\frac{J^2}{f_2(0,\eta)} \right).\label{geodesicaxx}
\eea

\subsection{Penrose limit}

In this section, we will perform the Penrose limit of 
the background in (\ref{mmmaaa}). 

To proceed, we need to assume 
that the functions in (\ref{functkk}) admit an expansion 
close to $\sigma=0$ of the form
\bea
& & f_1(\sigma,\eta)\sim \hat{f}_{1}(\eta)+ F_{1,2}(\eta)\sigma^2+O(\sigma^4),\;\;\; f_2(\sigma,\eta)\sim \hat{f}_{2}(\eta)
+ F_{2,2}(\eta)\sigma^2+O(\sigma^4),\nonumber\\[5pt]
& & f_3(\sigma,\eta)\sim \hat{f}_{3}(\eta)+ F_{3,2}(\eta)\sigma^2+O(\sigma^4),\;\;\; f_4(\sigma,\eta)\sim \hat{f}_{4}(\eta)
+ F_{4,2}(\eta)\sigma^2+O(\sigma^4).\nonumber\\[5pt]
& &  f_5(\sigma,\eta)\sim \hat{f}_{5}(\eta)+ F_{5,2}(\eta)\sigma^2+O(\sigma^4),\;\;\; f_6(\sigma,\eta)\sim a + F_{6,2}(\eta)\sigma^2
+O(\sigma^4),\nonumber\\[5pt]
& &  f_7(\sigma,\eta)\sim \hat{f}_{7}(\eta)+ F_{7,2}(\eta)\sigma^2+O(\sigma^4),\;\;\; \Delta(\sigma,\eta)\sim \hat{\Delta}_{4}(\eta)
+ \Delta_{2}(\eta)\sigma^2+O(\sigma^4).
\label{expansionsww}
\eea
Notice that the function $f_6(\sigma=0,\eta)=\hat{f}_6(\eta)=a$, 
is assumed to be a constant.
We will discuss the generality of the proposal and  assumptions below. 

Now, we scale the coordinates as in previous Sections 
\footnote{We use a slightly different gauge. The final configuration is the same.}, 
\bea
& & \chi=\frac{\pi}{2}+ \frac{z}{L},\;\;\; \rho_{AdS}=\frac{r}{2L},\;\;\; \sigma=\frac{x}{L},\nonumber\\[5pt]
& & d\eta= \frac{1}{\hat{f}_1 \sqrt{\hat{f}_3}} \sqrt{\frac{1}{4} -\frac{J^2}{\hat{f}_2}} du,\\[5pt]
& & d\xi= \frac{J}{\hat{f}_1\hat{f}_2} du +\frac{dw}{L},\;\;\;  dt=\frac{1}{4\hat{f}_1} du +\frac{J}{L} dw - \frac{dv}{L^2}.\nonumber
\label{scalinggmxx}
\eea
This produces a pp-wave  metric
\bea
& & ds^2= 2 du dv + \hat{f}_1\hat{f}_3\left(dx^2+\frac{\hat{f}_4}{\hat{f}_3} d\beta^2\right) 
+(\hat{f}_1 \hat{f}_2 -4 J^2 \hat{f}_1)dw^2+ \hat{f}_1\hat{f}_2 dz^2 +\hat{f}_1
(dr^2+r^2 d\Omega_3)+\nonumber\\[5pt]
& & - du^2\Big[  U_r r^2 + U_z z^2 + U_x x^2   \Big].\nonumber\\[5pt]
& & U_r=\frac{1}{16 \hat{f}_1},\;\;\;\; 
U_z= \frac{J^2}{\hat{f}_1\hat{f}_2},\;\;\; 
U_x= \left(\frac{4J^2-\hat{f}_2}{4\hat{f}_1\hat{f}_2\hat{f}_3}\right) F_{3,2} 
-J^2\frac{F_{2,2}}{\hat{f}_1\hat{f}_2^2}
\label{gmpprosen}
\eea
While the  rest of the Neveu-Schwarz and Ramond fields are, in the large-$L$ expansion,
\bea
& &B_2= \hat{f}_5 dz\wedge du    ,\;\;\;\;  \nonumber\\[5pt]
& & e^{2\Phi}= \frac{\hat{f}_1^3 \hat{f}_3}{\hat{\Delta}}\;\;\;\; F_2= 0,\;\;\;\; \nonumber\\[5pt]
& & F_4= H_1[u] du\wedge dx\wedge dz\wedge d\beta + 
H_2[u] du\wedge dw\wedge dz\wedge d\beta.
\eea
The functions $H_1, H_2$ are involved combinations of the $f_i$ defined above.

We can move the metric in (\ref{gmpprosen}) 
to Brinkmann coordinates as we indicated 
in Section \ref{section2xx}---see around (\ref{ds2ppNATD}),
\bea
& & ds^2=
2 du dv +\left(dx^2+\frac{\hat{f}_4}{\hat{f}_3} d\beta^2\right) +dw^2+  dz^2 +(dr^2+r^2 d\Omega_3)+\nonumber\\
& & - du^2\Big[  M^2_r r^2 + M^2_z z^2 + M^2_x x^2 +M^2_w w^2   \Big].\nonumber\\[5pt]
& & M^2_r=U_r+ \frac{(\hat{f}'_1)^2 -2 \hat{f}_1 \hat{f}_1'' }{4 \hat{f}_1^2},\nonumber\\[5pt]
& &  M_x=U_x+ \frac{1}{ (2\hat{f}_1\hat{f}_3)^2} \Big((\hat{f}_1\hat{f}_3')^2  + \hat{f}_3^2\big[(\hat{f}_1')^2 
-2 \hat{f}_1\hat{f}_1''  \big] -2 \hat{f}_1\hat{f}_3 \big[\hat{f}_1'\hat{f}_3'+\hat{f}_1\hat{f}_3''\big]  \Big) \nonumber\\[5pt]
& & 4 \hat{f}_1^2 (\hat{f}_2-4 J^2)^2 M_w=  (\hat{f}_2-4 J^2)^2 (\hat{f}_1')^2 -2 \hat{f}_1\hat{f}_1'\hat{f}_2' (\hat{f}_2-4 J^2)
- 2 \hat{f}_1' \hat{f}_1'' (\hat{f}_2-4 J^2)^2 +\nonumber\\[5pt]
& & (\hat{f}_1)^2 \Big[(\hat{f}_2)^2 + 2 \hat{f}_2''(\hat{f}_2-4 J^2)   \Big],\nonumber\\[5pt]
& &M_z=U_z+ \frac{1}{(2\hat{f}_1\hat{f}_2)^2} \Big((\hat{f}_1\hat{f}_2')^2  + \hat{f}_2^2\big[(\hat{f}_1')^2 
-2 \hat{f}_1\hat{f}_1''  \big] -2 \hat{f}_1\hat{f}_2 \big[\hat{f}_1'\hat{f}_2'+\hat{f}_1\hat{f}_2''\big]  \Big) \;,
\label{gmppbrinkmann}
\eea
where we have used the expressions for $U_x, U_z,U_r$ in (\ref{gmpprosen}).

If we specify for the particular potential function in (\ref{potentialdensityNATD}), we find the same masses we 
obtained for the non-Abelian pp-wave in (\ref{ds2ppNATD}).
Same goes for the potential in (\ref{potentialdensityTD}) and the pp-wave of Section \ref{section1xx}, 
that in the Abelian T-dual case was already in Brinkmann coordinates.
The general lesson this exercise teaches 
is that the masses of the pp-wave associated with Gaiotto-Maldacena backgrounds are dependent on the affine parameter. 
This, we believe, is an effect of the particular geodesic and expansion we are choosing.

As promised above, we now discuss the genericity of the 
expansion in (\ref{expansionsww}). The general solution 
to a Gaiotto-Maldacena problem, as defined in (\ref{ecuagm1})
was discussed in the papers 
\cite{ReidEdwards:2010qs}. Here, we will consider a particular
situation, where the solution is expanded for values small values 
$\sigma\sim 0$. We propose a potential function,
\be
V(\sigma,\eta)= F(\eta)+ a\eta \log\sigma 
+\sum_{k=1}^{\infty} h_k(\eta) \sigma^{2k}.
\label{solucionnuestra}
\ee
For this potential to satisfy the Laplace equation
(\ref{ecuagm1}) we impose
\bea
& & h_1(\eta)=-\frac{1}{4}F(\eta)'',\nonumber\\[5pt]
& & h_{k}(\eta)= -\frac{1}{4k^2} h_{k-1}(\eta)''.
\label{solucionrec}
\eea
Every function in the expansion is given in terms of 
derivatives of the function $F(\eta)$. For example, 
the Sfetsos-Thompson solution corresponds to $F(\eta)=\frac{\eta^3}{3}$.
Also,  notice that the charge density---see (\ref{ecuagm1})--  is
$\lambda= a\eta$ for any solution in this family. Hence the second boundary
condition is not satisfied (the quiver field theory is infinitely long).

It would be interesting 
to solve for the function $F(\eta)$, that makes the 
coefficients for the masses in (\ref{gmppbrinkmann}) the simplest possible.

\section{Geometric aspects of the pp-wave}

In this appendix we collect useful formulas for various geometric quantities of the pp-wave, such as the spin-connection, the 
Riemann tensor and the Ricci tensor. For a ten-dimensional pp-wave in Brinkmann coordinates with line element
\begin{equation}
\label{ppWaveFrame}
 ds^2 = 2 \, du \, dv + \sum\limits_{i = 1}^{8} (dy^i)^2 + \cH(u,y^i) \, du^2,
\end{equation}
one can define the orthonormal basis
\begin{equation}
 e^{-} = du \ , \qquad e^{+} = dv + \frac{1}{2} \, \cH \, du \ , \qquad e^{i} = dy^i \ ,
\end{equation}
so that
\begin{equation}
\label{ppWaveMetricFrame}
 ds^2 = 2 \, e^{+} e^{-} + \sum\limits_{i = 1}^{8} (e^i)^2 = \eta_{a b} \, e^a \, e^b \ ,
\end{equation}
where the non-vanishing components of $\eta_{ab}$ are $\eta_{+-} = \eta_{-+} = 1$ and $\eta_{ij} = \delta_{ij}$.

From the previous frame, one can easily compute the components of the spin-connection,
\begin{equation}
 \om_{-i} = - \om_{i-} = \om^{+i} = - \om^{i+} = \frac{1}{2} \, \partial_i \cH \, du \ .
\end{equation}
From the spin connection we find that the only non-vanishing components of the Riemann tensor are
\begin{equation}
 R_{- i - j} = - \frac{1}{2} \, \partial_i \partial_j \cH \ .
\end{equation}
Also, the Ricci tensor has only one non-trivial component, namely
\begin{equation}
 R_{--} = - \frac{1}{2} \, \partial^i \partial_i \cH \ .
\end{equation}

Notice that if we define
\begin{equation}
\label{pmFrame}
 e^{\pm} = \frac{e^9 \pm e^0}{\sqrt{2}}  \ ,
\end{equation}
then the line element \eqref{ppWaveMetricFrame} takes the orthogonal form, for signature $(- , + , \ldots , +)$,
\begin{equation}
 ds^2 = - (e^0)^2 + (e^1)^2 + \ldots + (e^9)^2 \ .
\end{equation}

\section{Penrose limit of the Janus solution}\label{janusappendix}

In this appendix we consider the Penrose limit of the Janus solution, in order to explore the origin of the flow in the radial coordinate, and of the 
possible tachyon, for strings moving in the corresponding pp wave.

\subsection{The solution}

The Janusian solution \cite{Bak:2003jk} is a solution of the type IIB supergravity whose metric consists of a five-dimensional part 
that is asymptotically $AdS_5$, and an internal part which is a five-dimensional sphere, $S^5$. 
This solution can be seen as a deformation of the well-known $AdS_5 \times S^5$ solution and in \emph{string frame} it takes the form
\begin{equation}
 \begin{aligned}
  ds^2 & = L^2 \, e^{\frac{\Phi}{2}} \, \Big[   f(\mu) \, \Big(   d\mu^2 + ds^2_{AdS_4}  \Big) + ds^2_{S^5}  \Big] \ ,
  \\[5pt]
  \Phi & = \Phi(\mu) \ ,
  \\[5pt]
  F_5 & = 4 \, L^4 \, f^{\frac{5}{2}} \, d\m \wedge \textrm{Vol}_{AdS_4}  - 4 \, L^4 \,  \textrm{Vol}_{S^5} \ ,
 \end{aligned}
\end{equation}
where $\textrm{Vol}_{AdS_4}$ and $ \textrm{Vol}_{S^5}$ are the volume forms of $AdS_4$ and $S^5$ respectively and they are oriented 
in such a way that $F_5$ is self-dual. For later convenience we choose the following parametrization for $AdS_4$ and $S^5$:
\begin{equation}
 ds^2_{AdS_4} = - \cosh^2 r \, dt^2 + \sinh^2 r \, d\Om^2_2 + dr^2 \ , \qquad ds^2_{S^5} = d\th^2 + \cos^2\th \, d\phi^2 + \sin^2\th \, d\Om^2_3 \ .
\end{equation}

The above ansatz is a solution of the type IIB supergravity provided that the dilaton $\Phi$ and the function $f$ satisfy the 
first order differential equations
\begin{equation}
\label{JanusianCond}
 \Phi' = \frac{c_0}{f^{\frac{3}{2}}} \ , \qquad f'^2 = 4 \, f^3 - 4 \, f^2 + \frac{c^2_0}{6 \, f} \ ,
\end{equation}
where $c_0$ is a constant and the prime stands for differentiation with respect to $\m$. Integrating, one obtains
\be
\mu=\int_{f_{\rm min}}^f\frac{d\tilde f}{2\sqrt{\tilde f^3-\tilde f^2+\frac{c_0}{24}\frac{1}{f}}}.
\ee

\subsection{Penrose limit for motion in $\m, \, \phi$ }

In order to compare with the Penrose limits for T-dual solutions considered in the main text, we need to consider a Penrose limit giving a pp wave 
for motion in $\mu$ and $\phi$, situated at $r = 0 , \, \th = 0$. 

The condition for a null geodesic for motion on $\mu$ is 
\be
g^{il}\d_l g_{\mu\mu}=0,\;\;\forall l\neq \mu,
\ee
which is indeed satisfied. For motion in $\phi$, we obtain 
\be
g^{\theta\theta}\d_\theta g_{\phi\phi}=0\Rightarrow \sin \theta \cos \theta=0\;,
\ee
which implies $\theta=0$ or $\theta=\pi/2$. The latter gives zero coefficient for $d\phi^2$, so we must put $\theta=0$ for this geodesic.

Considering a geodesic moving only in $\mu$, we expand around it as 
\be
\mu=x^+-\frac{x^-}{L^2},\;\;\;
t=x^+,\;\;\; r=\frac{\rho}{L},\;\;\; \a_a=\frac{y_a}{L},\; a=1,2,3, \;\; \theta_i=\frac{x_i}{L}\;, i=1,..,5.
\ee
Expanding and using $\cosh^2 r\simeq 1+\rho^2/L^2$, we get the metric
\be
ds^2=f(x^+)\left[-2dx^+dx^--\rho^2(dx^+)^2+\sum_{a=1}^3 dy_a^2\right]+\sum_{i=1}^5 dx_i^2\;,
\ee
the dilaton is now $\phi=\phi(x^+)$, and the 5-form is
\be
F_5=\frac{1}{L}\left[-8f(x^+)^{5/2}dx^+\wedge dx^-\wedge dy^1\wedge dy^2\wedge dy^3+2dx^1\wedge ... \wedge dx_5\right]\rightarrow 0.
\ee

We see that in this case, the overall function $f(x^+)$ flows in the same way as in the solution before the Penrose limit. Note however that the 
pp wave solution is not in Brinkmann form. 

Consider next the more general case of motion in both $\mu$ and $\phi$.

The Lagrangian for a null particle moving along the directions $t , \, \m , \, \phi$ is given by
\begin{equation}
L^{-2} \, \cL = e^{\frac{\Phi}{2}} \, \Big[   f \, \Big(  \dot{\m}^2  - \dot{t}^2  \Big) + \dot{\phi}^2  \Big] \ ,
\end{equation}
where dots stand for differentiation with respect to the 
affine parameter $u$.

The Lagrangean is cyclic in (independent of) $t$ and $\phi$, meaning that we have the conservation laws
\begin{equation}
 \frac{\partial \cL}{\partial \dot{t}} = p_t = \textrm{const} \ , \qquad \frac{\partial \cL}{\partial \dot{\phi}} = p_\phi = \textrm{const} \ .
\end{equation}
We fix 
\be
L^{-2}\frac{p_t}{2}=-e^{\phi/2}f \dot t=-1\;,\;\;\;
L^{-2}\frac{p_\phi}{2}=\dot \phi e^{\phi//2}=J.
\ee
Then the null condition, ${\cal L}=0$, gives
\begin{equation}
\label{eqmu}
 \dot{\m}^2 = \frac{e^{- \Phi}}{f^2} \, \Big(  1 - J^2 \, f  \Big) \ .
\end{equation}

Consider the expansion, near the null geodesic,
\be
 r=\frac{\bar r}{L},\;\;\; \a_a=\frac{y_a}{L},\; a=1,2,3, \;\; \theta_i=\frac{x_i}{L}\;, i=1,..,4\;,
\ee
where we have not rescaled $\phi$. For the coordinates along which the geodesic moves, we write the expansion
\begin{equation}
 \begin{aligned}
  d\m & = \frac{e^{- \frac{\Phi}{2}}}{f} \, \sqrt{1 - J^2 \, f} \, du -\frac{4e^{-\phi/2}}{f}J\sqrt{\frac{1-4J^2}{1-16J^2 e^{-\Phi}f^{-2}}}\frac{dw}{L}
  +\frac{4}{\sqrt{1-J^2f}}\frac{dv}{L^2}\ ,
  \\[10pt]
  dt & = \frac{e^{- \frac{\Phi}{2}}}{f} du \ ,
  \\[10pt]
  d\phi & = e^{- \frac{\Phi}{2}} \, J \, du +\sqrt{\frac{1-4J^2}{1-16J^2 e^{-\Phi}f^{-2}}} \frac{dw}{L}\ .
 \end{aligned}
\end{equation}

The resulting geometry is
\begin{equation}
 \begin{aligned}
  ds^2 = &  2 \, du \, dv +  e^{\frac{\Phi}{2}} \, f \, \Big(   d\bar{r}^2 + \bar{r}^2 \, d\Om^2_2  \Big) 
  + e^{\frac{\Phi}{2}} \, \Big(   dx^2 + x^2 \, d\Om^2_3  \Big) + e^{\frac{\Phi}{2}} \, \big(  1 - J^2 \, f  \big) \, dw^2
  \\[10pt]
  & - e^{-\frac{\Phi}{2}} \, \Bigg[  \frac{\bar{r}^2}{f} + J^2 \, x^2  \Bigg] \, du^2 \ .
 \end{aligned}
\end{equation}
Moreover, for the RR form we find
\begin{equation}
 F_5 = 4 \, J \, e^{-\frac{\Phi}{2}} \, \Big[  f^{\frac{3}{2}} \, \sqrt{1 - J^2 \, f} \, \bar{r}^2 \, du \wedge dw \wedge d\bar{r}
  \wedge d\Om_2 + x^3 \, du \wedge dx \wedge d\Om_3   \Big] \ ,
\end{equation}
where $d\Om_2$ and $d\Om_3$ are the volume forms of the two- and three-sphere respectively.

In order to verify the equations of motion of the type IIB supergravity, one has to take into account that the dilaton, $\Phi$, 
and the function $f$ must satisfy the conditions \eqref{JanusianCond}. Or, since $\mu$ is a function of $u$ through \eqref{eqmu},
\begin{equation}
\label{JanusianCondNew}
 \dot{\Phi} =  c_0 \, \frac{e^{- \frac{\Phi}{2}}}{f^{\frac{5}{2}}} \, \sqrt{1 - J^2 \, f} \ , \qquad \dot{f}^2 =  e^{- \Phi} \, 
 \Big(1 - J^2 \, f \Big) \, \Big(4 \, f - 4 + \frac{c^2_0}{6 \, f^3}\Big) \ ,
\end{equation}

\subsection{pp-wave in Brinkmann coordinates}

In order to bring the line element of the pp-wave found in the previous section into Brinkmann coordinates it is more convenient to write it as
\begin{equation}
\label{ppWaveJanus}
 \begin{aligned}
  ds^2 = &  2 \, du \, dv +  e^{\frac{\Phi}{2}} \, f \, \sum\limits_{i = 1}^3 \big( dy^i \big)^2+ e^{\frac{\Phi}{2}} \,  
  \sum\limits_{i = 1}^4 \big( dx^i \big)^2 + e^{\frac{\Phi}{2}} \, \big(  1 - J^2 \, f  \big) \, dw^2
  \\[5pt]
  & - \frac{e^{-\frac{\Phi}{2}}}{f} \, \Bigg[  \sum\limits_{i = 1}^3 \big( y^i \big)^2 + J^2 f \, \sum\limits_{i = 1}^4 \big( x^i \big)^2  \Bigg] \, du^2
 \end{aligned}
\end{equation}
and
\begin{equation}
\label{RRppWaveJanus}
 \begin{aligned}
   F_5 = 4 \, J \, e^{-\frac{\Phi}{2}} \, \Big[  & f^{\frac{3}{2}} \, \sqrt{1 - J^2 \, f} \, du \wedge dw \wedge dy^1 \wedge dy^2 \wedge dy^3 
   \\[10pt]
   & + du \wedge dx^1 \wedge dx^2 \wedge dx^3 \wedge dx^4   \Big] \ .
 \end{aligned}
\end{equation}

From the above expression for the line element we can read the coefficients of $\big( dx^i \big)^2$, $\big( dy^i\big)^2$ and $dw^2$,
\begin{equation}
  A_x = e^{\frac{\Phi}{2}} \ , \qquad A_y = f \, e^{\frac{\Phi}{2}} \ , \qquad A_w = e^{\frac{\Phi}{2}} \, \big(  1 - J^2 \, f  \big) \ ,
\end{equation}
where $A_x , \, A_y$ and $A_w$ are functions of $u$. Then one can apply the coordinate transformations (see also \eqref{BrinkmannTransf})
\begin{equation}
 \begin{aligned}
     & x^i \rightarrow \frac{x^i}{\sqrt{A_x}} \ , \qquad y^i \rightarrow \frac{y^i}{\sqrt{A_y}} \ , \qquad w \rightarrow \frac{w}{\sqrt{A_w}} \ ,
     \\[10pt]
     & v \rightarrow v + \frac{1}{4} \, \Bigg[   \frac{\dot{A_x}}{A_x} \, \sum\limits_{i = 1}^4 \big( x^i \big)^2 
     + \frac{\dot{A_y}}{A_y} \, \sum\limits_{i = 1}^3 \big( y^i \big)^2 + \frac{\dot{A_w}}{A_w} \, w^2  \Bigg] \ .
 \end{aligned}
\end{equation}
Under these transformations the line element \eqref{ppWaveJanus} becomes
\begin{equation}
 \begin{aligned}
  ds^2 = &  2 \, du \, dv +  \sum\limits_{i = 1}^3 \big( dy^i \big)^2+\sum\limits_{i = 1}^4 \big( dx^i \big)^2 + dw^2
  \\[5pt]
  & - \Bigg[  \Big(   \frac{e^{- \Phi}}{f^2} - \tilde{F}_y   \Big) \, \sum\limits_{i = 1}^3 \big( y^i \big)^2 
  + \Big(  J^2 e^{- \Phi} - \tilde{F}_x \Big) \, \sum\limits_{i = 1}^4 \big( x^i \big)^2 - \tilde{F}_w \, w^2  \Bigg] \, du^2 \ ,
 \end{aligned}
\end{equation}
where
\begin{equation}
 \tilde{F}_i = \frac{1}{4} \frac{\dot{A}^2_i}{A^2_i} + \frac{1}{2} \frac{d}{du} \Big(   \frac{\dot{A}_i}{A_i}  \Big) \ , \qquad i = x \ , y \ , w \ .
\end{equation}
Notice that the transformation applies also in $x^i$ and $y^i$ of the $du^2$ component of the metric. One easily finds
\begin{equation}
 \begin{aligned}
  \tilde{F}_x & = \frac{\ddot{\Phi}}{4} + \frac{\dot{\Phi}^2}{16} \ ,
  \\[5pt]
  \tilde{F}_y & = \frac{\dot{f} \, \dot{\Phi}}{4 \, f} + \frac{\ddot{f}}{2 \, f} + \frac{\ddot{\Phi}}{4} + \frac{\dot{\Phi}^2}{16} - \frac{\dot{f}^2}{4 \, f^2} \ ,
  \\[5pt]
  \tilde{F}_w & = \frac{1}{16} \, \Bigg[  4 \, \ddot{\Phi} + \dot{\Phi}^2 - \frac{4 \, J^2 \dot{f} \, \dot{\Phi}}{1 - J^2 f}
   - \frac{4 \, J^4 \dot{f}^2}{\big(  1 - J^2 f  \big)^2} - \frac{8 \, J^2 \ddot{f}}{1 - J^2 f}   \Bigg] \ . 
 \end{aligned}
\end{equation}

Similarly the RR form in \eqref{RRppWaveJanus} transforms as
\begin{equation}
 \begin{aligned}
   F_5 = 4 \, J \, e^{-\frac{\Phi}{2}} \, \Bigg[  & \frac{   f^{\frac{3}{2}} \, \sqrt{1 - J^2 \, f}}{A^{\frac{3}{2}}_y \, \sqrt{A_w}} \, du 
   \wedge dw \wedge dy^1 \wedge dy^2 \wedge dy^3 
   \\[5pt]
   & + \frac{1}{A^2_x} \, du \wedge dx^1 \wedge dx^2 \wedge dx^3 \wedge dx^4   \Bigg] \ .
 \end{aligned}
\end{equation}
To summarize, the pp-wave of the Janusian solution in Brinkmann form is
\begin{equation}
 \begin{aligned}
  ds^2 = &  2 \, du \, dv + \sum\limits_{i = 1}^3 \big( dy^i \big)^2+ \sum\limits_{i = 1}^4 \big( dx^i \big)^2 + dw^2  
  - \Bigg[  F_y \, \sum\limits_{i = 1}^3 \big( y^i \big)^2 + F_x \sum\limits_{i = 1}^4 \big( x^i \big)^2 + F_w \, w^2  \Bigg] \, du^2
  \\[5pt]
  F_5 = & 4 \, e^{- \frac{3 \, \Phi}{2}} \, J \, \Big(  du \wedge dw \wedge dy^1 \wedge dy^2 \wedge dy^3 
  + du \wedge dx^1 \wedge dx^2 \wedge dx^3 \wedge dx^4  \Big) \ ,
 \end{aligned}
\end{equation}
with
\begin{equation}
 \begin{aligned}
  F_x & = e^{- \Phi} \, J^2 - \frac{\ddot{\Phi}}{4} - \frac{\dot{\Phi}^2}{16} \ ,
  \\[5pt]
  F_y & = \frac{e^{-\Phi}}{f^2} + \frac{\dot{f}^2}{4 \, f^2} - \frac{\dot{f} \, \dot{\Phi}}{4 \, f} - \frac{\ddot{f}}{2 \, f} - \frac{\ddot{\Phi}}{4} - \frac{\dot{\Phi}^2}{16} \ ,
  \\[5pt]
  F_w & = \frac{1}{16} \, \Bigg[  \frac{4 \, J^2 \dot{f} \, \dot{\Phi}}{1 - J^2 f} + \frac{4 \, J^4 \dot{f}^2}{\big(  1 
  - J^2 f  \big)^2} + \frac{8 \, J^2 \ddot{f}}{1 - J^2 f} - 4 \, \ddot{\Phi} - \dot{\Phi}^2   \Bigg] \ .
 \end{aligned}
\end{equation}
Or, after implementing \eqref{JanusianCondNew},
\begin{equation}
 \begin{aligned}
  F_x & = e^{- \Phi} \, J^2 + c^2_0 \, \frac{e^{- \Phi}}{16 \, f^5} \, \Big(  1 - J^2 \, f  \Big) + c_0 \, \frac{e^{- \Phi}}{8 \, 
  \sqrt{6} \, f^5} \, \Big(   5 - 4 \, J^2 \, f  \Big) \, \sqrt{c^2_0 - 24 \, \big(   1 - f  \big) \, f^3} \ ,
  \\[5pt]
  F_y & = F_x + c^2_0 \, \frac{e^{- \Phi}}{24 \, f^5} \, \Big(   4 - 3 \, J^2 f  \Big) \ ,\qquad F_w = F_x - c^2_0 \, J^2 \, \frac{e^{- \Phi}}{8 \, f^4} \ .
 \end{aligned}
\end{equation}


\begin{thebibliography}{99}
\bibitem{tHooft:1973alw} 
  G.~'t Hooft,
  {\it A Planar Diagram Theory for Strong Interactions},
  Nucl.\ Phys.\ B {\bf 72}, 461 (1974).

\bibitem{Maldacena:1997re} 
  J.~M.~Maldacena,
  {\it The Large N limit of superconformal field theories and supergravity},
  Int.\ J.\ Theor.\ Phys.\  {\bf 38}, 1113 (1999),
  [Adv.\ Theor.\ Math.\ Phys.\  {\bf 2}, 231 (1998)],
  \href{http://arxiv.org/abs/hep-th/9711200}{{\tt hep-th/9711200}}.

\bibitem{Berenstein:2002jq} 
  D.~E.~Berenstein, J.~M.~Maldacena and H.~S.~Nastase,
  {\it Strings in flat space and pp waves from N=4 superYang-Mills},
  JHEP {\bf 0204}, 013 (2002),
  \href{http://arxiv.org/abs/hep-th/0202021}{{\tt hep-th/0202021}}.

\bibitem{Blau:2002dy}
  M.~Blau, J.~M.~Figueroa-O'Farrill, C.~Hull and G.~Papadopoulos,
  {\it Penrose limits and maximal supersymmetry},
  Class.\ Quant.\ Grav.\  {\bf 19} (2002) L87,
  \href{http://arxiv.org/abs/hep-th/0201081}{{\tt hep-th/0201081}}.
  
\bibitem{Blau:2002mw}
  M.~Blau, J.~M.~Figueroa-O'Farrill and G.~Papadopoulos,
  {\it Penrose limits, supergravity and brane dynamics},
  Class.\ Quant.\ Grav.\  {\bf 19} (2002) 4753,
  \href{http://arxiv.org/abs/hep-th/0202111}{{\tt hep-th/0202111}}.
  
\bibitem{Sadri:2003pr} 
  D.~Sadri and M.~M.~Sheikh-Jabbari,
  {\it The Plane wave / superYang-Mills duality},
  Rev.\ Mod.\ Phys.\  {\bf 76}, 853 (2004),
  \href{http://arxiv.org/abs/hep-th/0310119}{{\tt hep-th/0310119}}.

\bibitem{Horowitz:1989bv} 
  G.~T.~Horowitz and A.~R.~Steif,
  {\it Space-Time Singularities in String Theory},
  Phys.\ Rev.\ Lett.\  {\bf 64}, 260 (1990).


\bibitem{delaOssa:1992vci} 
  X.~C.~de la Ossa and F.~Quevedo,
  {\it Duality symmetries from nonAbelian isometries in string theory},
  Nucl.\ Phys.\ B {\bf 403}, 377 (1993),
  \href{http://arxiv.org/abs/hep-th/9210021}{{\tt hep-th/9210021}}.
  
\bibitem{Alvarez:1993qi} 
  E.~Alvarez, L.~Alvarez-Gaume, J.~L.~F.~Barbon and Y.~Lozano,
  {\it Some global aspects of duality in string theory},
  Nucl.\ Phys.\ B {\bf 415}, 71 (1994),
  \href{http://arxiv.org/abs/hep-th/9309039}{{\tt hep-th/9309039}}.
  E.~Alvarez, L.~Alvarez-Gaume and Y.~Lozano,
  {\it On nonAbelian duality},
  Nucl.\ Phys.\ B {\bf 424}, 155 (1994),
  \href{http://arxiv.org/abs/hep-th/9403155}{{\tt hep-th/9403155}}.
  E.~Alvarez, L.~Alvarez-Gaume and Y.~Lozano,
  {\it A Canonical approach to duality transformations},
  Phys.\ Lett.\ B {\bf 336}, 183 (1994),
  \href{http://arxiv.org/abs/hep-th/9406206}{{\tt hep-th/9406206}}.
  S.~Elitzur, A.~Giveon, E.~Rabinovici, A.~Schwimmer and G.~Veneziano,
  {\it Remarks on nonAbelian duality},
  Nucl.\ Phys.\ B {\bf 435}, 147 (1995),
  \href{http://arxiv.org/abs/hep-th/9409011}{{\tt hep-th/9409011}}.
  C.~Klimcik and P.~Severa,
  {\it Dual nonAbelian duality and the Drinfeld double},
  Phys.\ Lett.\ B {\bf 351}, 455 (1995),
  \href{http://arxiv.org/abs/hep-th/9502122}{{\tt hep-th/9502122}}.
  Y.~Lozano,
  {\it NonAbelian duality and canonical transformations},
  Phys.\ Lett.\ B {\bf 355}, 165 (1995),
  \href{http://arxiv.org/abs/hep-th/9503045}{{\tt hep-th/9503045}}.
  K.~Sfetsos, 
  {\it NonAbelian duality, parafermions and supersymmetry}, Phys.\ Rev. \ D {54}, 1682 (1996), 
   \href{http://arxiv.org/abs/hep-th/9602179}{{\tt hep-th/9602179}}.

  
 

\bibitem{Sfetsos:2010uq}
  K.~Sfetsos and D.~C.~Thompson,
  {\it On non-abelian T-dual geometries with Ramond fluxes},
  Nucl.\ Phys.\ B {\bf 846} (2011) 21,
  \href{http://arxiv.org/abs/arXiv:1012.1320}{{\tt arXiv:1012.1320}}.



\bibitem{Lozano:2012au}
  Y.~Lozano, E.~O Colgain, D.~Rodriguez-Gomez and K.~Sfetsos,
  {\it Supersymmetric $AdS_6$ via T Duality},
  Phys.\ Rev.\ Lett.\  {\bf 110} (2013) no.23,  231601,
  \href{http://arxiv.org/abs/arXiv:1212.1043}{{\tt arXiv:1212.1043}}.

\bibitem{Itsios:2013wd}
  G.~Itsios, C.~Nunez, K.~Sfetsos and D.~C.~Thompson,
  {\it Non-Abelian T-duality and the AdS/CFT correspondence:new N=1 backgrounds},
  Nucl.\ Phys.\ B {\bf 873} (2013) 1,
  \href{http://arxiv.org/abs/arXiv:1301.6755}{{\tt arXiv:1301.6755}}.
  G.~Itsios, C.~Nunez, K.~Sfetsos and D.~C.~Thompson,
  {\it On Non-Abelian T-Duality and new N=1 backgrounds},
  Phys.\ Lett.\ B {\bf 721} (2013) 342,
  \href{http://arxiv.org/abs/arXiv:1212.4840}{{\tt arXiv:1212.4840}}.

  


 \bibitem{variosa2}
   Y.~Lozano, E.~O Colgain, K.~Sfetsos and D.~C.~Thompson,
  {\it Non-abelian T-duality, Ramond Fields and Coset Geometries},
  JHEP {\bf 1106} (2011) 106,
  \href{http://arxiv.org/abs/arXiv:1104.5196}{{\tt arXiv:1104.5196}}.
  G.~Itsios, Y.~Lozano, E.~O Colgain and K.~Sfetsos,
  {\it Non-Abelian T-duality and consistent truncations in type-II supergravity},
  JHEP {\bf 1208} (2012) 132,
  \href{http://arxiv.org/abs/arXiv:1205.2274}{{\tt arXiv:1205.2274}}.
  J.~Jeong, O.~Kelekci and E.~O Colgain,
  {\it An alternative IIB embedding of F(4) gauged supergravity},
  JHEP {\bf 1305}, 079 (2013),
  \href{http://arxiv.org/abs/arXiv:1302.2105}{{\tt arXiv:1302.2105}}.
  A.~Barranco, J.~Gaillard, N.~T.~Macpherson, C.~Nunez and D.~C.~Thompson,
  {\it G-structures and Flavouring non-Abelian T-duality},
  JHEP {\bf 1308}, 018 (2013),
  \href{http://arxiv.org/abs/arXiv:1305.7229}{{\tt arXiv:1305.7229}}.
  N.~T.~Macpherson,
  {\it Non-Abelian T-duality, $G_2$-structure rotation and holographic duals of $N=1$ Chern-Simons theories},
  JHEP {\bf 1311}, 137 (2013),
  \href{http://arxiv.org/abs/arXiv:1310.1609}{{\tt arXiv:1310.1609}}.
  
  \bibitem{Lozano:2013oma} 
  Y.~Lozano, E.~O.~Colgain and D.~Rodriguez-Gomez,
  {\it Hints of 5d Fixed Point Theories from Non-Abelian T-duality},
  JHEP {\bf 1405}, 009 (2014),
  \href{http://arxiv.org/abs/arXiv:1311.4842}{{\tt arXiv:1311.4842}}.

  \bibitem{bbranch}
  J.~Gaillard, N.~T.~Macpherson, C.~Nunez and D.~C.~Thompson,
  {\it Dualising the Baryonic Branch: Dynamic SU(2) and confining backgrounds in IIA},
  Nucl.\ Phys.\ B {\bf 884}, 696 (2014),
  \href{http://arxiv.org/abs/arXiv:1312.4945}{{\tt arXiv:1312.4945}}.

\bibitem{variosb1} 
  D.~Elander, A.~F.~Faedo, C.~Hoyos, D.~Mateos and M.~Piai,
  {\it Multiscale confining dynamics from holographic RG flows},
  JHEP {\bf 1405}, 003 (2014),
  \href{http://arxiv.org/abs/arXiv:1312.7160}{{\tt arXiv:1312.7160}}.
  S.~Zacarias,
  {\it Semiclassical strings and Non-Abelian T-duality},
  Phys.\ Lett.\ B {\bf 737}, 90 (2014),
  \href{http://arxiv.org/abs/arXiv:1401.7618}{{\tt arXiv:1401.7618}}.
  E.~Caceres, N.~T.~Macpherson and C.~Nunez,
  {\it New Type IIB Backgrounds and Aspects of Their Field Theory Duals},
  JHEP {\bf 1408}, 107 (2014),
  \href{http://arxiv.org/abs/arXiv:1402.3294}{{\tt arXiv:1402.3294}}.
  P.~M.~Pradhan,
  {\it Oscillating Strings and Non-Abelian T-dual Klebanov-Witten Background},
  Phys.\ Rev.\ D {\bf 90}, no. 4, 046003 (2014),
  \href{http://arxiv.org/abs/arXiv:1406.2152}{{\tt arXiv:1406.2152}}.
  
\bibitem{Lozano:2014ata}
  Y.~Lozano and N.~T.~Macpherson,
  {\it A new AdS$_{4}$/CFT$_{3}$ dual with extended SUSY and a spectral flow},
  JHEP {\bf 1411} (2014) 115,
  \href{http://arxiv.org/abs/arXiv:1408.0912}{{\tt arXiv:1408.0912}}.
  
  
  \bibitem{variosa3}
  K.~Sfetsos and D.~C.~Thompson,
  {\it New ${\cal N} = 1$ supersymmetric $AdS_5$ backgrounds in Type IIA supergravity},
  JHEP {\bf 1411}, 006 (2014),
  \href{http://arxiv.org/abs/arXiv:1408.6545}{{\tt arXiv:1408.6545}}.
  O.~Kelekci, Y.~Lozano, N.~T.~Macpherson and E.~O.~Colgain,
  {\it Supersymmetry and non-Abelian T-duality in type II supergravity},
  Class.\ Quant.\ Grav.\  {\bf 32} (2015) no.3,  035014,
  \href{http://arxiv.org/abs/arXiv:1409.7406}{{\tt arXiv:1409.7406}}.
  
  \bibitem{Macpherson:2014eza} 
  N.~T.~Macpherson, C.~Nunez, L.~A.~Pando Zayas, V.~G.~J.~Rodgers and C.~A.~Whiting,
  {\it Type IIB supergravity solutions with AdS$_{5}$ from Abelian and non-Abelian T dualities},
  JHEP {\bf 1502}, 040 (2015),
  \href{http://arxiv.org/abs/arXiv:1410.2650}{{\tt arXiv:1410.2650}}.

\bibitem{variosa4}
   K.~S.~Kooner and S.~Zacarias,
  {\it Non-Abelian T-Dualizing the Resolved Conifold with Regular and Fractional D3-Branes},
  JHEP {\bf 1508}, 143 (2015),
  \href{http://arxiv.org/abs/arXiv:1411.7433}{{\tt arXiv:1411.7433}}.
  T.~R.~Araujo and H.~Nastase,
  {\it $\mathcal{N}=1$ SUSY backgrounds with an AdS factor from non-Abelian T duality},
  Phys.\ Rev.\ D {\bf 91}, no. 12, 126015 (2015),
  \href{http://arxiv.org/abs/arXiv:1503.00553}{{\tt arXiv:1503.00553}}.
  
\bibitem{Bea:2015fja}
  Y.~Bea, J.~D.~Edelstein, G.~Itsios, K.~S.~Kooner, C.~Nunez, D.~Schofield and J.~A.~Sierra-Garcia,
  {\it Compactifications of the Klebanov-Witten CFT and new AdS$_{3}$ backgrounds},
  JHEP {\bf 1505} (2015) 062,
\href{http://arxiv.org/abs/arXiv:1503.07527}{{\tt arXiv:1503.07527}}.  
  
  
  \bibitem{Lozano:2015bra} 
  Y.~Lozano, N.~T.~Macpherson, J.~Montero and E.~O.~Colgain,
  {\it New $AdS_3 \times S^2$ T-duals with $ \mathcal{N}=\left(0,4\right) $ supersymmetry},
  JHEP {\bf 1508}, 121 (2015),
  \href{http://arxiv.org/abs/arXiv:1507.02659}{{\tt arXiv:1507.02659}}.
    

   \bibitem{Lozano:2015cra} 
  Y.~Lozano, N.~T.~Macpherson and J.~Montero,
  {\it A $ \mathcal{N}=2 $ supersymmetric AdS$_{4}$ solution in M-theory with purely magnetic flux},
  JHEP {\bf 1510} (2015) 004,
  \href{http://arxiv.org/abs/arXiv:1507.02660}{{\tt arXiv:1507.02660}}.
%

\bibitem{Araujo:2015dba} 
  T.~R.~Araujo and H.~Nastase,
  {\it Non-Abelian T-duality for nonrelativistic holographic duals},
  JHEP {\bf 1511}, 203 (2015),
  \href{http://arxiv.org/abs/arXiv:1508.06568}{{\tt arXiv:1508.06568}}.
  
\bibitem{Macpherson:2015tka}
  N.~T.~Macpherson, C.~Nunez, D.~C.~Thompson and S.~Zacarias,
  {\it Holographic Flows in non-Abelian T-dual Geometries},
  JHEP {\bf 1511} (2015) 212,
  \href{http://arxiv.org/abs/arXiv:1509.04286}{{\tt arXiv:1509.04286}}.
  
  \bibitem{variosa5}
 L.~A.~P.~Zayas, V.~G.~J.~Rodgers and C.~A.~Whiting,
  {\it Supergravity solutions with AdS$_{4}$ from non-Abelian T-dualities},
  JHEP {\bf 1602}, 061 (2016),
  \href{http://arxiv.org/abs/arXiv:1511.05991}{{\tt arXiv:1511.05991}}.
 L.~A.~Pando Zayas, D.~Tsimpis and C.~A.~Whiting,
  {\it A Supersymmetric IIB Background with $AdS_4$ from Massive IIA},
  \href{https://arxiv.org/abs/1701.01643}{{arXiv:1701.01643}}.



\bibitem{Lozano:2016kum}
  Y.~Lozano and C.~N\'u\~nez,
  {\it Field theory aspects of non-Abelian T-duality and $ \mathcal{N}  =$ 2 linear quivers},
  JHEP {\bf 1605} (2016) 107,
  \href{http://arxiv.org/abs/arXiv:1603.04440}{{\tt arXiv:1603.04440}}.
  


\bibitem{completions}
  Y.~Lozano, N.~T.~Macpherson, J.~Montero and C.~Nunez,
  {\it Three-dimensional $ \mathcal{N}=4 $ linear quivers and non-Abelian T-duals},
  JHEP {\bf 1611}, 133 (2016),
  \href{http://arxiv.org/abs/arXiv:1609.09061}{{\tt arXiv:1609.09061}}.
  Y.~Lozano, C.~Nunez and S.~Zacarias,
  {\it BMN Vacua, Superstars and Non-Abelian T-duality},
  JHEP {\bf 1709}, 000 (2017),
  \href{http://arxiv.org/abs/arXiv:1703.00417}{{\tt arXiv:1703.00417}}.
  G.~Itsios, Y.~Lozano, J.~Montero and C.~Nunez,
  {\it The AdS$_{5}$ non-Abelian T-dual of Klebanov-Witten as a $ \mathcal{N}=1 $ linear quiver from M5-branes},
  JHEP {\bf 1709}, 038 (2017),
  \href{http://arxiv.org/abs/arXiv:1705.09661}{{\tt arXiv:1705.09661}}.




\bibitem{Corrado:2002wi} 
  R.~Corrado, N.~Halmagyi, K.~D.~Kennaway and N.~P.~Warner,
  {\it Penrose limits of RG fixed points and PP waves with background fluxes},
  Adv.\ Theor.\ Math.\ Phys.\  {\bf 6}, 597 (2003),
  \href{http://arxiv.org/abs/hep-th/0205314}{{\tt hep-th/0205314}}.

\bibitem{Gimon:2002sf} 
  E.~G.~Gimon, L.~A.~Pando Zayas and J.~Sonnenschein,
  {\it Penrose limits and RG flows},
  JHEP {\bf 0209}, 044 (2002)
  \href{http://arxiv.org/abs/hep-th/0206033}{{\tt hep-th/0206033}}.



\bibitem{Russo:2002rq} 
  Arkady Tseytlin private communication. See also J.~G.~Russo and A.~A.~Tseytlin,
  {\it On solvable models of type 2B superstring in NS NS and RR plane wave backgrounds},
  JHEP {\bf 0204}, 021 (2002),
  \href{http://arxiv.org/abs/hep-th/0202179}{{\tt hep-th/0202179}}.




  




\bibitem{Alishahiha:2002ev} 
  M.~Alishahiha and M.~M.~Sheikh-Jabbari,
  {\it The pp wave limits of orbifolded AdS(5) x S**5},
  Phys.\ Lett.\ B {\bf 535}, 328 (2002),
  \href{http://arxiv.org/abs/hep-th/0203018}{{\tt hep-th/0203018}}.


\bibitem{Mukhi:2002ck} 
  S.~Mukhi, M.~Rangamani and E.~P.~Verlinde,
  {\it Strings from quivers, membranes from moose},
  JHEP {\bf 0205}, 023 (2002),
  \href{http://arxiv.org/abs/hep-th/0204147}{{\tt hep-th/0204147}}.

\bibitem{Kuperstein:2003jd} 
  S.~Kuperstein,
  {\it On Penrose limit of elliptic branes},
  JHEP {\bf 0306}, 046 (2003),
  \href{http://arxiv.org/abs/hep-th/0303042}{{\tt hep-th/0303042}}.


  
  


\bibitem{Polchinski:1998rq} 
  J.~Polchinski,
  {\it String theory. Vol. 1: An introduction to the bosonic string},

  J.~Polchinski,
  {\it String theory. Vol. 2: Superstring theory and beyond}

\bibitem{Gaiotto:2009gz} 
  D.~Gaiotto and J.~Maldacena,
  {\it The Gravity duals of N=2 superconformal field theories},
  JHEP {\bf 1210}, 189 (2012),
  \href{http://arxiv.org/abs/arXiv:0904.4466}{{\tt arXiv:0904.4466}}.
  
  
\bibitem{vanGorsel:2017goj} 
  J.~van Gorsel and S.~Zacarias,
  {\it A Type IIB Matrix Model via non-Abelian T-dualities},
  \href{http://arxiv.org/abs/arXiv:1711.03419}{{\tt arXiv:1711.03419}}.

\bibitem{ArkaniHamed:2001ca} 
  N.~Arkani-Hamed, A.~G.~Cohen and H.~Georgi,
  {\it (De)constructing dimensions},
  Phys.\ Rev.\ Lett.\  {\bf 86}, 4757 (2001)
  \href{http://arxiv.org/abs/hep-th/0104005}{{\tt hep-th/0104005}}.

\bibitem{ArkaniHamed:2001nc} 
  N.~Arkani-Hamed, A.~G.~Cohen and H.~Georgi,
  {\it Electroweak symmetry breaking from dimensional deconstruction},
  Phys.\ Lett.\ B {\bf 513}, 232 (2001),
  \href{http://arxiv.org/abs/hep-ph/0105239}{{\tt hep-ph/0105239}}.


\bibitem{Maldacena:1998im} 
  J.~M.~Maldacena,
  {\it Wilson loops in large N field theories},
  Phys.\ Rev.\ Lett.\  {\bf 80}, 4859 (1998),
  \href{http://arxiv.org/abs/hep-th/9803002}{{\tt hep-th/9803002}}.





  
  
\bibitem{ReidEdwards:2010qs} 
  R.~A.~Reid-Edwards and B.~Stefanski, jr.,
  {\it On Type IIA geometries dual to N = 2 SCFTs},
  Nucl.\ Phys.\ B {\bf 849}, 549 (2011),
  \href{http://arxiv.org/abs/arXiv:1011.0216}{{\tt arXiv:1011.0216}}.
  O.~Aharony, L.~Berdichevsky and M.~Berkooz,
  {\it 4d N=2 superconformal linear quivers with type IIA duals},
  JHEP {\bf 1208}, 131 (2012),
  \href{http://arxiv.org/abs/arXiv:1206.5916}{{\tt arXiv:1206.5916}}.




\bibitem{Bak:2003jk}
  D.~Bak, M.~Gutperle and S.~Hirano,
  {\it A Dilatonic deformation of AdS(5) and its field theory dual},
  JHEP {\bf 0305} (2003) 072,
  \href{http://arxiv.org/abs/hep-th/0304129}{{\tt hep-th/0304129}}.



\end{thebibliography}
\end{document}